\documentclass[11pt,authoryear]{elsarticle}
\usepackage{arydshln}

\makeatletter
\def\ps@pprintTitle{%
 \let\@oddhead\@empty
 \let\@evenhead\@empty
 \def\@oddfoot{}%
 \let\@evenfoot\@oddfoot}
\makeatother

\usepackage{ulem}
\usepackage[margin=0.8in]{geometry}
\usepackage{setspace}

\usepackage{caption} 
\usepackage{subcaption} 
\usepackage[justification=centering]{caption}
\PassOptionsToPackage{hyphens}{url}
\usepackage{natbib}
\makeatother
\usepackage{amsthm,amsmath,amsfonts,amssymb,bm}
\usepackage{graphicx,psfrag,epsf}
\usepackage{epigraph,xcolor}
\usepackage{enumerate}
\usepackage{url}
\usepackage{booktabs}
\usepackage{multirow}
\usepackage{hyperref}
\usepackage{float}
\hypersetup{colorlinks=true,citecolor=blue,
	pdfpagemode=FullScreen,linktocpage=true}
\usepackage{cleveref}

\newcommand{\cor}{\mathrm{Corr}}

\newcommand{\logit}{\mathrm{logit}}

\newcommand{\ind}{\mathbb{I}}

\renewcommand{\L}{\mathcal{L}}

\newcommand{\RNum}[1]{\uppercase\expandafter{\romannumeral #1\relax}}
\DeclareMathOperator*{\argmin}{arg\,min}
\newcommand{\abs}[1]{\left\lvert #1 \right\rvert}
\newcommand{\norm}[1]{\left\| #1 \right\|}

\numberwithin{equation}{section}  

\begin{document}

\begin{frontmatter}

\title{A review and recommendations on variable selection methods in regression models for binary data}

\author{Souvik Bag}
\author{Kapil Gupta}
\author{Soudeep Deb}
  
\address{Indian Institute of Management Bangalore, Bannerghatta Main Road, Bangalore, KA 560076, India.}

\begin{abstract}
The selection of essential variables in logistic regression is vital because of its extensive use in medical studies, finance, economics and related fields. In this paper, we explore four main typologies (test-based, penalty-based, screening-based, and tree-based) of frequentist variable selection methods in logistic regression setup. Primary objective of this work is to give a comprehensive overview of the existing literature for practitioners. Underlying assumptions and theory, along with the specifics of their implementations, are detailed as well. Next, we conduct a thorough simulation study to explore the performances of sixteen different methods in terms of variable selection, estimation of coefficients, prediction accuracy  as well as time complexity under various settings. We take low, moderate and high dimensional setups and consider different correlation structures for the covariates. A real-life application, using a high-dimensional gene expression data, is also included in this study to further understand the efficacy and consistency of the methods. Finally, based on our findings in the simulated data and in the real data, we provide recommendations for practitioners on the choice of variable selection methods under various contexts.
\end{abstract}


\end{frontmatter}


\newpage

\section{Introduction} 
\label{sec:introduction}

With the advancement of data collection mechanisms, there is a surge of enormous datasets in recent years. In the context of regression models, it typically introduces a large number of predictors, all of which may not be necessary to analyze or predict the outcome appropriately. In fact, large number of variables often increase the complexity without necessarily inducing a substantial improvement in the fit or the prediction accuracy. Naturally, it is preferable to model a data with a suitably chosen smaller set of predictors which allows better model interpretability, and in some cases, superior prediction accuracy. That is one of the first and foremost reasons why variable selection is essential. Our focus in this paper is specifically on the logistic regression models (LRM), which is one of the most used regression and classification algorithms for binary data. We aim to provide a comprehensive review of existing classical variable selection methods in such models. 

The necessity of feature selection in LRM appears in a plethora of real-life applications from bioinformatics, medical research, finance, sports and other related fields. These research problems commonly consist of hundreds of predictors all of which may not be incorporated into the modeling structure. Gene expression data are the most popular examples in this regard (see \cite{shevade2003simple}, \cite{dara2017rough}, \cite{yang2018robust} for instance). In another related study, \cite{algamal2015penalized} focused on finding a smaller set of genes out of thousands that contribute the most in correctly classifying a tumor to be benign or malignant. Along a similar line, many medical studies (e.g.\ \cite{liang2013sparse}, \cite{bertoncelli2020predictmed}) aim to identify which factors out of hundreds increase the chance of a serious disease such as cerebral palsy, cancer, heart attack etc. In a finance-related application, \cite{tian2015variable} used variable selection in logistic regression to model bankruptcy based on important micro-economic factors. \cite{costa2017physical}, on the other hand, analyzed the effects of important factors behind activity-related injuries in children and adolescents. Such applications are abound in literature, and they establish why variable selection methods are crucial in modeling binary data, especially for high dimensional cases where number of predictors is more than the number of sample observations.

The research on variable selection methods started around the 1960s. Since then, it has been rapidly growing. There has been a substantial amount of work in both linear and generalized linear models over the years. However, unlike linear models which have been reviewed thoroughly by several authors (e.g.\ \cite{fan2010selective}, \cite{heinze2018variable}, \cite{desboulets2018review}, \cite{epprecht2021variable}), to the best of our knowledge, a detailed review on variable selection methods in logistic regression setup has not been done till date. The study by \cite{sanchez2018comparison} is perhaps the only work that includes considerable number of tree-based variable selection methods and compares them against three other popular approaches in a high-dimensional case. \cite{speiser2019comparison} did a limited study on the efficacy of different tree-based techniques used for predictive modeling. In another earlier paper, \cite{zellner2004variable} compared the performances of stepwise selection procedures with the bagging method of \cite{sauerbrei1999use}. Most recently, the performance of Smoothly Clipped Absolute Deviation (SCAD) and Adaptive Least Absolute Shrinkage and Selection operator (ALASSO) were evaluated by \cite{arayeshgari2020application} on a psychiatric distress dataset. 

Evidently, there is an immense need of an extensive review of existing variable selection methods in LRM. To that end, our contribution in this paper is three-fold. First, we provide an in-depth discussion of many classical procedures which can be clearly categorized in four different types. Implementation details of these methods are also presented as necessary. Second, we conduct simulation studies in low, moderate and high-dimensional cases to understand the performance of specific algorithms under different assumptions. A high-dimensional real-life dataset is also used in this aspect. Finally, acknowledging the fact that logistic regression is often used by machine learning practitioners as a classification problem for binary data, we evaluate both the inferential and the classification or prediction accuracy of all the methods in our study. We strongly believe that this comparative analysis would be helpful for a diverse class of practitioners. 

Rest of the paper is structured in the following way. In \Cref{sec:methods}, first we lay out the framework of the LRM and discuss the typology of variable selection procedures. Descriptions of different methods are provided next. Then, in \Cref{sec:simulation}, a detailed simulation study is presented to show the efficacy and comparison of sixteen methods. As a real-life application, we use a high-dimensional dataset from statistical genetics and discuss the results in \Cref{sec:real-analysis}. Finally, some recommendations and important concluding remarks are listed in \Cref{sec:conclusion}.

\section{Variable selection methods}
\label{sec:methods}

\subsection{Setup and typology of procedures}
\label{subsec:setup}

Before delving into the details of the variable selection methods, it is necessary to recall the structure of the LRM. In this setup, the response variable is always binary. If the outcome is a ``success'' (respectively, ``failure''), we assign 1 (respectively, 0) as the value of the response variable. Let us use $\boldsymbol Y=(y_1,y_2,\hdots,y_n)$ to denote the sample of response observations, which are assumed to depend on $m$ number of covariates. For the $i^{th}$ sample observation, the vector of covariates is denoted by $\boldsymbol{x}_i=(x_{i1},x_{i2},\hdots,x_{im})^\top$. The corresponding regression coefficients in the LRM are going to be denoted by $\boldsymbol{\beta} = (\beta_1,\beta_2,\hdots,\beta_m)^\top$, and that is our primary parameter of interest. We shall use $\boldsymbol{X}=[\boldsymbol{x}_1:\hdots:\boldsymbol{x}_n]^\top$ to denote the set of explanatory variables in the model. The order of the design matrix $\boldsymbol{X}$ is $n \times m$. Throughout this paper, $n$ and $m$ denote the number of observations and the total number of covariates respectively. Then, the LRM can be written using vector-matrix notations as 
\begin{equation}
\label{eqn:logistic}
    \logit (\boldsymbol Y \mid \boldsymbol{X}) = \boldsymbol{X} \boldsymbol{\beta},
\end{equation}
where $\logit(\boldsymbol Y \mid \boldsymbol{X})$ is used to denote the vector of logit transformation of $y_i$ given $\boldsymbol{x}_i$ for $1\leqslant i \leqslant n$, which is defined as
\begin{equation}
\label{eqn:logit}
    \logit (y_i\mid \boldsymbol{x}_i) = \log \frac{P(y_i=1 \mid \boldsymbol{x}_i)}{P(y_i=0 \mid \boldsymbol{x}_i)}. 
\end{equation}

Writing the probability $P(y_i = 1 \mid \boldsymbol{x}_i)$ as $\pi_i$, we note that it can be expressed as $\pi_i = \frac{\exp{\left(\boldsymbol{x}_i^\top \boldsymbol{\beta}\right)}}{1 + \exp{\left(\boldsymbol{x}_i^\top \boldsymbol{\beta}\right)}}$. Since the complete likelihood for the binary data is
\begin{equation}
    L(\boldsymbol{\beta}) = \prod_{i=1}^n \pi_i^{y_i}(1-\pi_i)^{1-y_i},
\end{equation}
the log likelihood for the regression coefficients can be written as
\begin{equation}
\log L(\boldsymbol{\beta}) = \sum _{i=1} ^{n} \left[ y_i \left( \boldsymbol{x}_i^\top \boldsymbol{\beta}\right) - \log\left(1+ \exp{\left(\boldsymbol{x}_i^\top \boldsymbol{\beta}\right)}\right)\right].
\end{equation}

The method of iteratively reweighted least squares (IRLS) is typically used to find the maximum likelihood estimate of $\boldsymbol{\beta}$ in LRM (\cite{green1984iteratively}). However, the standard procedure would suffer if there is an issue of multicollinearity among the predictors. Especially, when $m$ is large, it automatically increases the chances of both multicollinearity and overfitting. And in case of $m \geqslant n$, LRM fails to give a unique solution. Due to such issues, variable selection methods are needed.

In this paper, we are going to discuss the following four types of variable selection techniques:
\begin{itemize}
    \item Test-based methods,
    \item Penalty-based methods,
    \item Screening-based methods,
    \item Tree-based methods.
\end{itemize}

Arguably the most famous technique used by practitioners is the step-wise regression, which is a test-based method and was possibly the very first attempt at variable selection (\cite{breaux1967stepwise}). In such test-based approaches, each viable combination of input variables is considered to be a potential true model, and the selection is done based on some appropriate statistical measure such as $p$-values, adjusted $R^2$, Akaike Information Criterion (\cite{akaike1998information}), Bayesian Information Criterion (\cite{schwarz1978}), Mallow's $C_p$ (\cite{mallows1995}) etc. In the crudest possible way, such methods need to evaluate $2^m$ number of models if there are $m$ number of regressors. Consequently, even for $m\geqslant 15$, searching for the best model among these is computationally very expensive, and they become inconsistent in variable selection. Step-wise regression improves the computational efficiency to some extent, but by making a hard selection on every step, it makes choices that are locally optimal in each step but are suboptimal in general. The reader is referred to \cite{hurvich1990}, \cite{steyerberg1999stepwise} and \cite{whittingham2006we} for more detailed criticism on variable selection inconsistency of this type of procedures. Because of the aforementioned shortcomings and considering the challenges in high-dimensional data, we are going to exclude these methods from our simulation study and are going to focus on the other three typologies.

Moving on to the second type, it is worth mention that the penalty-based methods are perhaps the most discussed techniques in the last decade. Following the seminal work on Least Absolute Shrinkage and Selection operator (LASSO) by \cite{tibshirani1996regression}, these methods have been immensely popularized because of their nice properties in variable selection mechanism. In this typology of procedures, appropriate constraints are imposed on the regression coefficients' values through a penalty function. This introduces bias to reduce the variance. Penalty-based methods achieve sparsity by shrinking the near to zero values of the regression coefficients to precisely zero, thereby suggesting that those independent variables do not have any effect on the model. It is imperative to point out that the structure of the penalty functions has a significant impact on the number of variables selected and the amount of error in the regression coefficients. These methods are described in more detail in \Cref{subsec:penalty-methods}.

In case of screening-based methods, first a small number of variables are screened from the large number of predictors and then a smaller number of variables are selected using some penalty-based method or importance criteria. These methods are not particularly designed to do selection intrinsically but by ranking the importance of the variables. Screening-based methods are especially useful in an ultra high-dimensional setting where the number of variables grows with the number of observations (i.e.\ $m >> n$). \Cref{subsec:screeninng-methods} presents these methods in more detail.

One of the major disadvantages of the previous categories is that they heavily depend on the regularization parameters, which are not fixed and are calculated using cross-validation. Thus, they may fail to produce accurate and consistent results at times. In such situations, the fourth type of procedures i.e.\ the tree-based methods can be more useful. These methods utilize feature elimination characteristic embedded in random forests to select variables. In this work, we focus on six different tree-based approaches which are described in more detail in \Cref{subsec:tree-based}.

In \Cref{table:all}, we list the variable selection methods from each of the four types. Corresponding references and suitable R packages (if available) are also listed there. Note that our focus for the rest of the paper would be on the sixteen methods of the last three categories. It is also worth mention that there are several Bayesian methods and machine learning techniques that address the same problems, but we do not include them in this study and focus entirely on the aforementioned typologies.

\begin{table}[!ht]
\centering
\caption{Variable selection methods along with relevant references and R packages.}
{\small
\label{table:all}
\begin{tabular}{llllll}
\toprule
Type                           & { Name}                               & Reference & R-package                       \\ \midrule
\multirow{3}{*}{Test-based}    & Forward / Backward / Step-wise Selection                     & \cite{breaux1967stepwise}          & MASS      \\ 
\cmidrule(l){2-4}
                              & GETS                                &  \cite{hendry1999improving}                                      &     gets      \\ 
\cmidrule(l){2-4} 
                              & Autometrics                           & \cite{doornik2009autometrics}          & gets      \\ \midrule
\multirow{7}{*}{Penalty-based} & LASSO                                 & \cite{tibshirani1996regression}           & glmnet    \\ \cmidrule(l){2-4} 
                               & ElasticNet                           & \cite{zou2005regularization}          & glmnet    \\ \cmidrule(l){2-4} 
                               & Adaptive LASSO (ALASSO)                        & \cite{zou2006adaptive}                 & glmnet    \\ \cmidrule(l){2-4} 
                               & SparseStep                           & \cite{van2017SparseStep}                      & L0Learn   \\ \cmidrule(l){2-4} 
                               & Best Subset                           & \cite{hazimeh2020fast}          & L0Learn   \\  \cmidrule(l){2-4}
                               
                               & Smoothly Clipped Absolute Deviation (SCAD)   & \cite{fan2001variable}         & ncvreg    \\ \cmidrule(l){2-4} 
                               & Minimax Concave Penalty (MCP)             & \cite{zhang2010nearly}          & ncvreg    \\ \midrule
\multirow{3}{*}{Screening-based}                &  Sure Independence Screening (SIS)          & \cite{fan2008sure}              & SIS       \\ \cmidrule(l){2-4}
                               & Iterative Sure Independence Screening (ISIS) & \cite{fan2008sure}                                & SIS          \\ \cmidrule(l){2-4}
                               & Stable Iterative Variable Selection (SIVS)  & \cite{mahmoudian2021stable}          & SIVS      \\ \midrule
\multirow{6}{*}{Tree-based}    & Variable selection from random forests (VarSelRF)                                  & \cite{diaz2005variable}           & varSelRF          \\ \cmidrule(l){2-4} 
                               
                               & permutation importance (PIMP)       & \cite{altmann2010permutation}   & vita   \\ \cmidrule(l){2-4}
& Boruta                                  & \cite{kursa2010feature}           & Boruta          \\  \cmidrule(l){2-4}
                               & Variable selection using random forests (VSURF) & \cite{genuer2010variable}          &  VUSRF         \\
                               \cmidrule(l){2-4}
                               & Guided Regularized random forest (RRF)               &  \cite{deng2012feature}         &  RRF         \\ 
                               \cmidrule(l){2-4}
                               & The novel test approach (NTA) & \cite{janitza2018computationally} & vita \\
                               \hline
\end{tabular}
}
\end{table}

\subsection{Penalty-based methods}
\label{subsec:penalty-methods}

There are mainly two types of penalties that are imposed on the objective function of variable selection problems. They are norm-based penalties and concave penalties. We present these in the two subsections below. 

\subsubsection{Norm based penalties}
\label{subsubsec:norm-based}

Vector norms are required to describe these models. Recall that the $\L_d$ norm $(d \geqslant 1)$ for the $n$-dimensional vector $\bm q=(q_1,q_2,\hdots,q_n)^\top$ is defined as
\begin{equation}
    \label{eqn:vector-norm}
    \norm{\bm q}_d = \left(\sum_{i=1}^n \abs{q_i}^d\right)^{1/d}.
\end{equation}

$\L_0$ norm is also defined, but in actuality it is not a norm. It is a cardinality function to represent the total number of non-zero elements in a vector. It is defined as
\begin{equation}
    \label{eqn:zero-norm}
    \norm{\bm q}_0 = \#\left\{1 \leqslant i \leqslant n \mid q_i \neq 0 \right\},
\end{equation}
where $\#$ is used to denote the cardinality of a set.

In this class of methods that leverage various norm-based penalties, the objective is to find the set of regression coefficients by minimizing an objective function that combines the negative log-likelihood and a constraint imposed on the regression coefficients. In general, these methods work around the following minimization problem:
\begin{equation}
\label{eq:normeq}
  \hat{\bm\beta} = \argmin_{\bm\beta} \{ - \log L(\bm{\beta})+ \lambda\norm{\bm\beta}_\gamma ^2\}.
\end{equation}


Clearly, different degrees of the norm function would correspond to different algorithms. The three most popular methods, which take the form of \cref{eq:normeq}, are as follows.
\begin{equation}
    \gamma = \begin{cases}
    0 & \text{if it is SparseStep}, \\
    1 & \text{if it is LASSO}, \\
    2 & \text{if it is Ridge}.
    \end{cases}
\end{equation}

We should recall that \cite{frank1993statistical} earlier proposed a more general bridge estimator where $\gamma$ can be any nonnegative quantity, although an analytical solution was not developed in that work for any $\gamma$. Further note that the ridge regression (\cite{hoerl1970ridge}), which relies on the $\L_2$ penalty function, has an analytical solution, but is not sparse. So, it does not select variables but only does shrinkage. More generally, one can say that the use of $\L_\gamma$ norm selects variables when $0\leqslant \gamma \leqslant 1$, but for $\gamma > 1$, it only shrinks. In a very recent paper by \cite{wu2021can}, ridge selection operator (RSO) and adaptive version of RSO (ARSO) were developed to propose a way to select variables through ridge regression for usual linear models. Its extension in the generalized linear model (GLM) seems possible, but a precise formulation is yet to be obtained, primarily due to the fact that the ridge penalized solution does not have a closed-form expression for GLM. In light of the above, the case of $\gamma>1$ is not going to be a part of our discussion hereafter. 

Turning attention to the SparseStep method (\cite{van2017SparseStep}), observe that the $\L_0$ norm directly penalizes the number of nonzero coefficients and not their values, thereby inducing a high computational cost to search over the entire space. Subsequently, it makes SparseStep infeasible compared to others. LASSO regression is computationally faster and relies on the $\L_1$ norm to select variables by shrinking small coefficients exactly to zero. However, the results are biased even for large values of the coefficients (\cite{zou2006adaptive}). Moreover, if some of the variables are correlated, i.e.\ if they form a group, LASSO tends to select only one. And for high-dimensional problems where $m > n$, this method tends to select at most $n$ variables, which is a major concern in practice (\cite{efron2004least}). To circumvent these limitations, \cite{zou2005regularization} proposed the method of ElasticNet, which is essentially a linear combination of ridge penalty and LASSO penalty. In this approach, the $\L_1$ penalty generates a sparse model, while the quadratic penalty removes the limitation of selecting at most $n$ variables, encourages grouping effect and stabilizes the $\L_1$ regularization path. It also makes the loss function strongly convex and hence produces a unique estimate given by
\begin{equation}
  \hat{\bm\beta}_{\mathrm{ElasticNet}} =\argmin_{\bm\beta}  \left \{ - \log L(\bm\beta)+ \lambda_1\norm{\beta}_1 + \lambda_2\norm{\bm\beta}_2^2 \right\}.
\end{equation}

Another problem with LASSO, as pointed out by \cite{meinshausen2006high}, is that it tends to select noise variables even for an optimally chosen tuning parameter ($\lambda$). \cite{zou2006adaptive} resolved it by developing an Adaptive LASSO (ALASSO) algorithm that enjoys the oracle property by utilizing the adaptive weights given to different coefficients. If $\bm{x}\circ\bm{y}$ represents the element-wise Hadamard product of two vectors $\bm{x}$ and $\bm{y}$, then the ALASSO estimate is given by
\begin{equation}
    \hat{\bm\beta}_{\mathrm{ALASSO}} = \argmin_{\bm\beta} \left\{-\log L(\bm\beta)+ \lambda \norm{\bm{w \circ \beta}}_1 \right\}.
\end{equation}

A pivotal issue with the above estimation problem is the initial choice of $\bm{w}$. Adaptive LASSO originally used MLE estimates as initial weights, but the assumption is invalid for high-dimensional data. \cite{buhlmann2011statistics} used LASSO estimates as initial weights, but since LASSO itself is biased and would produce zero coefficients for some variables, using it as weights may not be a perfect choice. \cite{zou2006adaptive} suggested using some unbiased estimate $\hat{\bm\beta}$ so that the ALASSO estimates can achieve oracle properties. As oracle property is an asymptotic guarantee when $n \to \infty$, it may not hold for a small sample size. In practice, the weights need not be exact, and one can get reasonable values of $\bm w$ from an initial estimate of $\bm\beta$, possibly from ordinary least squares, LASSO or ridge.


The aforementioned methods have been naturally extended to the GLM case. Refer to the work by \cite{friedman2010regularization} who formulated the theoretical development of a regularized path for such models via coordinate descent so that LASSO, ALASSO and ElasticNet can be applied in the LRM setup.  An exciting application of ALASSO in LRM can be found in \cite{cui2021adaptive} where they used this approach to select an important set of covariates for early diagnosis of Alzheimer's disease.

We close this subsection with another method that resolves an important limitation of the methods discussed above. Note that the above approaches either penalize the model size i.e.\ the number of non-zero coefficients (for example, SparseStep) or the size of the coefficients (for example, LASSO, ElasticNet etc.), but not both at the same time. Recently, \cite{hazimeh2020fast} proposed an extended family of $\L_0$ based estimators that are further regularized by $\L_q$ norm to avoid overfitting issues in both low and high signal-to-noise phenomena. They developed a fast algorithm to perform these sparse regularizations using coordinate descent and local combinatorial optimization algorithms. This approach is called the `Best Subset' method and uses the estimate
\begin{equation}
\hat{\bm{\beta}} = \arg \min_{\bm{\beta}} \{-\log L(\bm\beta) + \lambda_0\norm{\bm\beta}_0 + \lambda_q\norm{\bm\beta}^q_q  \},
\end{equation}
where $q \in \{1,2\}$ depends on the type of desired additional regularization.

\subsubsection{Concave penalties}
\label{subsubsec: concave-based}

In the second type of penalty-based methods, an appropriately chosen concave function $p_\lambda(\cdot)$ is used which leads to the general framework 
\begin{equation}
\label{eq:concave-penalty-general}
    \hat{\bm\beta} =\argmin_{\bm\beta} \left \{  -\log L(\bm\beta)+ p_{\lambda}(\bm\beta) \right\}.
\end{equation} 

Non-negative garrotte by \cite{breiman1995better} was one of the first methods to use non-convex penalties. If $f_+$ denotes the non-negative part of $f$, i.e.\ the quantity $\max\{f,0\}$, then the concave penalty used in non-negative garrotte is given by 
\begin{equation}
    p_{\lambda}(\bm\beta) = n\lambda \sum_{j = 1}^{m}\left(1-\frac{\lambda}{\beta_j^2}\right)_+.
\end{equation}

Due to variable selection inconsistency, the above method was rapidly disused. \cite{fan2001variable} then proposed the Smoothly Clipped Absolute Deviation (SCAD) method and it is arguably the most prominent procedure in this typology. Recall that LASSO shrinks all the least square estimates by an identical amount of $\lambda/2$. If their absolute values are less than $\lambda/2$, they are shrunk to precisely zero. Because of this phenomenon, LASSO induces bias in estimating the larger values of beta estimates. SCAD penalty addresses this issue. For lower values of $\hat{\beta}_j$, the penalty is the same as LASSO, and it shrinks the coefficients to zero. For the value of $\hat{\beta}_j$ between $\lambda$ and $\gamma \lambda$, the penalty function smoothly transitions to quadratic and gradually relaxes the penalization rate. The penalty then remains constant for all the values of $\hat{\beta}_j$ larger than $\gamma \lambda$. Thus, in SCAD, \cref{eq:concave-penalty-general} uses
\begin{equation}
\label{eq: cocave-loglikelihood}
p_{\lambda}(\bm\beta) = \sum_{j=1}^m p(\abs{\beta_j};\lambda), 
\end{equation}
where, $p(t;\lambda)$ is the SCAD penalty indexed by a regularization parameter $\lambda \geqslant 0$ and is given by
\begin{equation}
\label{eq:scad-penalty}
   p(t;\lambda) = \lambda \int_{0}^t \mathrm{min}\left\{1 , \frac{(\gamma - x/\lambda)_+}{(\gamma - 1)} \right\}dx
\end{equation}
for some $\gamma > 2$. In practice, one can choose different $\lambda$ values for different $\beta_j$ but we shall take the conventional approach of imposing same penalty on each regression coefficient.

Another concave penalty similar to SCAD is the Minimax Concave Penalty (MCP), proposed by \cite{zhang2010nearly}. MCP also starts with the same penalization rate as LASSO but smoothly relaxes it to zero as the values of the regression coefficients increase. In contrast, in SCAD, this rate remains constant for a while and then decreases to zero. The penalty function corresponding to MCP is of the form of \cref{eq: cocave-loglikelihood}, where 
\begin{equation}
\label{eq:mcp-penalty}
    p(t; \lambda) = \lambda \int_0^{t} \left (1 - \frac{ x}{\gamma\lambda} \right )_+ dx,
\end{equation}
for appropriately chosen parameters $\lambda$ and $\gamma$.

The tuning Parameter $\gamma$ for SCAD or MCP (would be different for the two methods) controls the concavity of the penalty function. As $\gamma \to \infty$, both penalties converge to the $\L_1$ penalty, and the bias is minimised when $\gamma$ is minimum. Still, the solution becomes unstable because there may exist multiple local minima of the loss function. Two applications of concave penalty-based methods in logistic regression can be found in \cite{rezapur2020risk} and \cite{yan2011sparse}, whereas a comparative study between the two approaches can be found in \cite{zhang2007penalized}.

\subsection{Screening-based methods}
\label{subsec:screeninng-methods}

As mentioned in \Cref{subsec:setup}, penalty-based methods may fail in the case of ultra high dimensional data. \cite{fan2008sure} developed the Sure Independence Screening (SIS) method for such scenarios. Sure screening means that for large amount of data, after variable selection, the probability of all important variables surviving converges to 1. It filters out variables that have a weak correlation with the response. To put it mathematically, let $M_* = \{1 \leqslant i \leqslant m : \beta_i \neq 0 \}$ be the true sparse model and let the non-sparsity size be $s= |M_*|$. Other $(m-s)$ variables can be correlated with the response via some linkage to the important variables in the model. If $\bm w=\bm{X}^\top \bm Y$ is a vector obtained by the component-wise regression where the matrix $\bm{X}$ and the vector $\bm Y$ are standardized such that each variable has mean 0 and variance 1, then $\bm w$ represents a vector of marginal correlations. Next, for $\nu \in (0,1)$, arrange the elements of $\bm w$, i.e.\ the marginal correlations, in ascending order of magnitudes to get the sub-model 
$$M_\nu = \{1 \leqslant i \leqslant m : |w_i| \text{ is among the first } [\nu n] \text{ largest of all } w_i\}.$$ 

It is a variable selection method based on correlation ranking of size $d = [\nu n]<n$. Observe that the computational cost is minimal, as this approach is just about multiplying a matrix of order $m\times n$ with a vector of $n \times 1$ and finding the largest [$n\nu$] elements in it. After applying SIS to reduce the dimension from $m$ to much smaller $d$, SCAD, ALASSO or any other variable selection technique can be used for further selection and estimation. The parameter $d$ is typically selected depending on the following algorithm to be performed after SIS. Overall, this method speeds up variable selection drastically and improves estimation accuracy in high-dimensional settings. 

A caveat, however, is that SIS may fail to select some important variables. To overcome this problem, Iterative Sure Independence Screening (ISIS) has been introduced where a large scale variable screening is applied before a careful variable selection. Here, an SIS-based model selection is used first to select a primary set of variables, say $A_1$. Next, the residuals from regressing the response $Y$ on the chosen set of variables in $A_1$ are treated as new responses and applied the same method as previous. The advantage of this iterative procedure is that, because the residuals are uncorrelated with the selected variables in $A_1$, it significantly weakens the possibility of choosing an unimportant variable that has a high correlation with $Y$ through the chosen variables.

Stable iterative Variable Selection (SIVS), proposed by \cite{mahmoudian2021stable}, is another screening-based method for variable selection. SIVS works in five steps. It starts with removing the redundant features and standardizing the columns. Then, a predefined number of models are created using cross-validation. Based on the results, variable importance score (VIMP) is calculated such that the variables selected by most models and those who have a major contribution in predicting the outcome get a higher score. The variables with a VIMP score of zero are directly eliminated in the next step. Other variables with low VIMP scores are also eliminated following a suitable criteria discussed in the original paper.   

Before moving on to the tree-based methods, we want to briefly mention two other methods which are beneficial for variable selection in linear models but are not yet completely developed for the GLM case. First of the two, Covariance Assisted Screening and Estimation (CASE) (\cite{ke2014covariance}) is a two-step screening-based method that deals with the case where the signals (non-zero coordinates of $\bm \beta$) are sparse as well as weak (absolute value of the non-zero coefficients are small). The two steps are patching and screening (PS) and patching and estimation (PE). In the PS step, a sequential $\chi^2$ test is used to look for candidates in each signal island of a graph of strong dependence. In the PE step, penalized likelihood is used to re-investigate each candidate in the hope to solve the problem of false positives. The problem with CASE is to find an appropriate filtering method to sparsify the non-sparse variance-covariance matrix.

\cite{ke2017covariate} proposed another method for variable ranking, known as Factor Adjusted Covariate Assisted Ranking (FA-CAR). In the FA step, the authors advocated for the use of principal component analysis (PCA) to sparsify the covariance matrix when the variables are strongly correlated. The CAR step exploits the sparse covariance matrix for the ranking of variables. Note that the concept of using PCA to sparsify the covariance matrix is similar to CASE, where suitable linear filtering was used, but the two methods have a different objective. FA-CAR's primary goal is to rank the variables, which are crucial in many statistical analyses. Variable selection can be done via appropriate thresholds as a by-product of the variable rankings.

\subsection{Tree-based methods}
\label{subsec:tree-based}

Recall that the determination of the regularization parameters in penalty-based or screening-based methods is done through cross-validation, which sometimes results in inconsistency in variable selection. In such cases, tree-based methods are beneficial as they do not rely upon any such parameter and achieve good results even in the presence of missing data, outliers etc. Let us start this section with a brief background on decision trees as they are the building blocks of various tree-based methods. 

The decision tree is a supervised machine learning algorithm with the tree's structure. Each internal node represents different features, each branch exhibits a decision rule, and each leaf represents an outcome. Random forest (RF) is an ensemble of decision trees (\cite{breiman2001random}), and the outcome depends on the decision tree outcomes through a majority rule in case of classification. Feature selection is an inherent part of random forest and is carried out by ranking the features according to some importance scores. Because these scores alone may not be sufficient for identifying the most significant features, multiple types of variable selection methods are available, depending on how a RF grows.

One of the earliest variable selection method to utilize random forests is VarSelRF, proposed by \cite{diaz2005variable}. This variable selection method was developed with the primary goal of selecting variables in gene expression data for biomedical research. The main objective of VarSelRF is to find the minimal number of genes that can deliver good predictive performance in clinical settings while avoiding redundant factors. When the values of a variable are permuted randomly in a tree node, one of the most reliable measures of variable importance in a RF model is the drop in classification accuracy. VarSelRF generates a number of random forests iteratively using this measure of variable importance, deleting the variables with the lowest importance scores at each iteration. Random Forest delivers the out-of-bag error (OOBE) for each fitted tree based on the OOB samples (observations not used in the RF construction) as a measure of error. The smallest set of variables with OOBE within the range of $u$ standard deviation from the minimal error rate of all forests is chosen after measuring the OOBE for each RF. Clearly, $u = 0$ corresponds to the set of variables with the lowest OOBE. When $u = 1$, it is known as the ``1 s.e.\ rule'', and it allows us to choose the least number of genes with an OOBE that is not too far from the minimum OOBE. It is worth noting that, in order to avoid overfitting, bootstrap sampling methods are utilized in VarSelRF to assess prediction error. For more in-depth discussions, the reader is referred to \cite{saffari2009line}. 

Next, we look at Boruta (\cite{kursa2010feature}), which is one of the most popular RF-based variable selection methods. By comparing its importance scores with randomly shuffled original features termed shadow variables, Boruta gives a criterion for picking essential features. The basic notion behind this method is that it introduces randomness into the system by introducing randomly shuffled shadow attributes, resulting in random correlations between shadow attributes and decision attributes. The extended dataset is then used to build a RF, and the relevance scores of all attributes are recorded. If an attribute has a greater significance score than the highest importance score earned by shadow variables in a particular random forest run, it is considered essential for that run. The shadow variables are permuted at random for each new RF iteration. A hit is defined as the number of times an attribute has a higher relevance score than maximal importance of random attributes. Using the properties of binomial distribution with probability of success $p = 0.5$, it is easy to argue that for $N$ number of RF runs, the expected number of hits is $E(N) = 0.5N$, while the standard deviation is $0.25N$. Now, a variable is deemed important (respectively insignificant) if it has a significantly higher (respectively lower) number of hits than expected. The method comes to a halt when only the most important variables remain in the test or when the maximum number of iterations with some uncertain attributes has been reached. As an attractive application, Boruta was used by \cite{naik2019optimal} to determine the best feature set for a stock prediction challenge. Meanwhile, \cite{hallmark2020developing} leveraged this technique to work on roadway safety models in an LRM setup.
 
Variable selection using random forest (VSURF) (\cite{genuer2010variable})  is another popular algorithm that selects variables using the mechanism embedded in the random forest. It employs a two-step procedure. In the first stage, the features are ranked according to their Variable Importance (VI) score, which for a variable $X^j$ is defined as
\begin{equation*}
    \mathrm{VI}(X^j) = \frac{1}{n}  \sum_t (\widehat{\mathrm{OOBE}}^j_t -  \mathrm{OOBE}_t),
\end{equation*}
where $\mathrm{OOBE}_t$ denotes the error of a single tree $t$, out of total $n$ number of trees. When the values of $X^j$ are randomly permuted we get the perturbed sample and corresponding error is denoted by $\widehat{\mathrm{OOBE}}^j_t$. Features with a lower VI score are then removed. When compared to useless factors, significant variables have a lot more variation in VI ratings. As a result, the threshold value is calculated using the standard deviation of the unnecessary variables' VI scores.

The next phase is variable selection, which involves choosing two subsets of variables. The interpretation step is where all the variables that are relevant to the outcome variable are chosen, even if there is a lot of repetition. With the initial $k$ number of variables, and for $k = 1$ to $m$, a hierarchical collection of RF models is created. We choose the variables in the model, say $m'$, that result in the lowest OOBE. The prediction step comes next, in which a small number of factors are chosen with the goal of predicting the response variable. A subset of variables is obtained for the prediction stage by building an ascending number of random forests and executing a test criteria for picking variables in a stepwise way. A variable will be selected only if the error decreases more than a threshold value, or in other words, if the added variable contributes to a considerable reduction in OOBE over the average variation achieved by adding noisy variables. The mean of absolute value of first order differentiated OOBE with variables $m'$ and $m$ determines the threshold. Mathematically, if $\mathrm{OOBE}(j)$ is the out-of-bag error of RF model with $j$ variables, then the threshold is
\begin{equation*}
    \frac{1}{m-m'} \sum_{m'}^{m-1} \abs{\mathrm{OOBE}(j+1) - \mathrm{OOBE}(j)}.
\end{equation*}

VSURF has been employed in a variety of medical studies. For example, \cite{ganggayah2019predicting} used it to uncover critical parameters affecting breast cancer survival rates, while \cite{yin2019predicting} applied it to predict endometrioid endometrial adenocarcinoma disease prognosis using gene expression and clinical trial data. \cite{virdi2019feature} combined VSURF and LASSO to pick a small set of variables and model mechanical features of investment casting, which is an interesting application of distinct flavor.


We now turn attention to the regularized random forest (RRF), developed by \cite{deng2012feature} in an attempt to pick features more effectively with the use of a regularization step over a general random forest. Let $\mathrm{gain}(X_j)$ be the measured information gain for each variable $X_j$ at each node, and if a variable has maximum gain, it is chosen to divide the tree at each node. Let $F$ be the feature set used in the previous steps to split the tree. When the final tree is constructed, $F$ becomes the final feature set that has been chosen. The objective behind the RRF framework is to avoid picking a new feature $X_j$ until the gain is significantly higher than $\max \{\mathrm{gain}(X_j)\}$ for all $X_i \in F$. Thus, for a regularization parameter $\lambda \in [0,1]$, a new gain measure is defined as
\begin{equation*}
    \mathrm{gain}(X_j) = \begin{cases}
    \lambda \times \mathrm{gain}(X_j) \ ; \ X_j \notin F \\
    \mathrm{gain}(X_j) \ ; \ X_i \in F,
    \end{cases}
\end{equation*}

A regularized tree is one that is formed by using the above splitting process. The features in the $F$ set are the features that have been chosen. Because of the selection mechanism, numerous features may have the same information gain at leaf nodes, and insignificant variables may be incorrectly picked for small sample numbers. \cite{deng2013gene} developed an enhanced version of the RRF approach dubbed the guided RRF, which uses relevance ratings from regular random forests to guide feature selection in the RRF algorithm. \cite{adam2017detecting} used this technique for feature selection and classification in early detection of phaeosphaeria leaf spot infestations in maize fields in a binary data context. \cite{sylvester2018applications} employed these methods to choose key panels from large scale nucleotide polymorphisms (SNP) panels for fine-scale population assignment.

Interestingly, the RF models are biased in the direction of categorical variables with a large number of categories. To that end, \cite{altmann2010permutation} developed a heuristic, called permutation importance (PIMP), for normalizing feature importance metrics that can correct the bias. In a non-informative environment, PIMP is based on repeated permutations of the outcome vector to estimate the distribution of assessed importance for each variable. The observed importance's $p$-value serves as a corrected measure of feature relevance. The method was carried out in two stages. For each variable, an arbitrary variable importance measure is computed in the first step. The outcome variable is then randomly permuted (say $s$ times) to disrupt its relationship with all of the predictor variables in the second step. After that, the data from each permutation is utilized to build a RF and assess the variable importance for all of the predictors. As a result, all variables have $s$ significance measures, which can be thought of as realizations from an unknown null distribution. These values are then used to calculate the empirical $p$-values, which are the percentages of importance scores higher than the initial score. The $p$-values are thereafter utilized to pick variables that are relevant. A crucial caveat of using PIMP is that, despite rectifying some flaws with RF-based techniques and being tractable in low-dimensional scenarios, it becomes computationally infeasible when the number of variables is very large.


Random forests are used in all of the tree-based methods outlined above for classification and variable importance measures. These metrics assign ratings to variables based on their importance. The lack of a natural cutoff to distinguish between important and unimportant variables is a disadvantage though. Several approaches, such as those based on hypothesis testing (\cite{hapfelmeier2013new}) and PIMP, have been developed to overcome this. However, these are computationally intensive and require the computation of random forests multiple times. \cite{janitza2018computationally} improved in that aspect and developed a computationally efficient method for determining variable relevance. This approach is known as the new testing method (NTA), and it determines whether a predictor variable significantly improves the trees' predictive performance. The authors demonstrated that in simulations, albeit limited to classification problems, NTA successfully identified at least as many important predictor variables as the PIMP while preserving the type-I error. Below, we take a quick look at this method.

Suppose in a classification problem, $(X_1, X_2,\hdots, X_p)$ are the $p$ covariates and $f(\cdot)$ and $Y$ are the predicted and observed class taking values in $\{1,2,\hdots,k\}$. If $X_j^*$ is a random replication of $X_j$ unaffected by the outcome variable or any other predictor variables (see \cite{gregorutti2017correlation}), then the importance score for $X_j$ is defined by
\begin{equation*}
    \mathrm{VI}_j = P\left(Y \neq f(X_1,\hdots, X_j^*,\hdots, X_p)\right) - P\left(Y \neq f(X_1,\hdots, X_j,\hdots, X_p)\right).
\end{equation*}

If the value of $X_j$ is not reliant on the outcome, permuting it will have no effect and $\mathrm{VI}_j$ should be zero. If $X_j$ is dependent on the outcome,  $P(Y \neq f( X_1,\hdots, X_j^*,\hdots, X_p))$ is expected to be greater than $P(Y \neq f(X_1,\hdots, X_j,\hdots, X_p))$ and in that case $\mathrm{VI}_j$ will be greater than zero. With this in view, \cite{janitza2018computationally} presented NTA with the null and alternative hypotheses as
\begin{equation*}
    H_0 :  \mathrm{VI}_j \leqslant 0 \  \mathrm{against} \ H_1 :  \mathrm{VI}_j > 0.
\end{equation*}

Here, the null distribution is calculated using the importance scores of non-important variables, i.e.\ variables with zero or negative scores. It should be noted that the null distribution is constructed using a cross-validated variant of permutation important measurements in order to produce a smooth and symmetric (around zero) distribution curve, which may not be possible with traditional permutation significance scores. Then, to choose relevant variables, $p$-values related to variable importance to predictor variables are employed. An application of NTA in LRM can be found in \cite{sun2020lung}, who used this method to detect a set of genes important in lung cancer diagnosis.

\section{Simulation study}
\label{sec:simulation}

\subsection{Setups and implementation details}

In our simulation study, we aim to compare the variable selection parsimony and predictive accuracy for the methods mentioned in the previous section. Eighteen different setups are considered to mimic a wide range of problems encountered in real life. Throughout this section, keeping up with the earlier notations, we use $n$, $m$ and $p$ to denote the number of sample observations, the total number of covariates and the number of important covariates, respectively. The simulation frameworks are summarized in \Cref{tab:simulation-setting} and relevant discussions are provided next.

\begin{table}[!ht]
\caption{Simulation setups used in this paper. Here, $n$ is the sample size, $m$ is the total number of covariates, $p$ is the number of important covariates, and $X_i$ stands for the $i^{th}$ covariate.}
\label{tab:simulation-setting}
\centering
\begin{tabular}{cccc}
\toprule
$n$ & $m$ & $p$ & Correlation between covariates \\ \midrule
\multirow{3}{*}{(100, 200, 500)} & 10   & 3  & \multirow{3}{*}{Independent}   \\ 
                     & 100  & 5 &   \\ 
                     & 1000 & 10 &    \\ \midrule
\multirow{3}{*}{(100, 200, 500)} & 10   & 3  & \multirow{3}{*}{$\cor(X_i,X_j)=0.5^{\abs{i-j}}$}  \\ 
                     & 100  & 5 &  \\ 
                     & 1000 & 10 & \\ \bottomrule
\end{tabular}
\end{table}
We follow the usual logistic regression framework as in \cref{eqn:logistic}. First step is to generate synthetic observations for $m$ number of covariates ($X_i$) to constitute the $n\times m$ data matrix $\bm X$. To that end, two different correlation structures are utilized in our simulation. In the first case, we use an ideal assumption of independence across all covariates. In the second case, following \cite{hazimeh2020fast}, an exponential hierarchical correlation structure is adopted. In particular, we assume that the correlation between the $i^{th}$ and the $j^{th}$ regressor is of the form $\rho^{\abs{i-j}}$. For this simulation, we shall use $\rho = 0.5$ in all experiments. This indicates moderate correlation between nearby covariates, and negligible correlation amongst far apart covariates. Now, each row of the design matrix is simulated as a random observation from an appropriately chosen (based on desired correlation structure) $m$-dimensional zero-mean multivariate normal distribution. 

For each correlation structure, we simulate $n \in \{100, 200, 500\}$ observations with the number of predictor variables $m$ varying in the set $\{10,100,1000\}$ respectively for the three choices of $n$. Observe that these choices represent low dimensional, moderate dimensional and high dimensional data, respectively. In these three cases, the number of important variables are chosen as $p \in \{3, 5, 10\}$, respectively. Now, to simulate the $m\times 1$ parameter vector $\boldsymbol{\beta}$, we randomly pick the $p$ indexes that represent the important regressors. These coefficients are generated using the formula $Z+(0.5)\ind\{Z>0\}-(0.5)\ind\{Z\leqslant 0\}$, where $Z$ is a random observation from a normal distribution. Note that it ensures the condition $\abs{\beta_j} \geqslant 0.5$ for all important coefficients $\beta_j$, and thereby avoids the cases of weak signals. Other $(m-p)$ coefficients are set to be zero and imply unimportant variables. 

To assess and compare the performances of the competing variable selection methods, we repeat each experiment 100 times and evaluate the average accuracy across those repetitions. 

A brief account of the R implementation details for all the methods is warranted at this point. We use the R package ``glmnet'' by \cite{friedman2010regularization} to perform LASSO, ALASSO and ElasticNet. The first two methods are performed in default settings. The weights used for regression coefficients $(\boldsymbol{\beta})$ in ALASSO are taken as reciprocals of the absolute values of their ridge regression estimates. For ElasticNet, we use five-fold cross-validation technique to find the optimal values of $\alpha$ and $\lambda$ parameters using the ``caret'' package (\cite{caret}). The ``L0Learn'' package (\cite{l0learn}) is used to implement the SparseStep and Best Subset methods. In both situations, a five-fold cross-validation technique is used to identify the optimal values of the $\lambda$ and $\gamma$ parameters, after which the $L_0$ penalty is used to find the sparse estimates and a combination of the $L_0$ and $L_2$ penalties is used to get the Best Subset estimates. In the implementation of both approaches, the maximum support size at which the regularization path will be ended is set to 20. The methods that rely on concave penalties are executed using the R package ``ncvreg'' (\cite{ncvreg}). The default setting is used to implement these methods. Here, the $\lambda$ parameter is determined using ten-fold cross-validation. 

Turn attention to the screening-based methods next. For SIS and ISIS, we use the default settings in the ``SIS'' package (\cite{sis_package}). SIVS is carried out using the ``sivs'' package (\cite{mahmoudian2021stable}). To screen important variables, the ``strictness'' argument is set to be 0.5, as in our settings, the number of important variables is assumed to be quite low. For real data analysis, we can try out different strictness levels and choose what is optimal. As suggested by the authors in the corresponding vignette, this argument is dependent on the problem at hand.

We utilize ``randomForest'' (\cite{randomForest}) to grow the trees in all of the tree-based approaches. The ``varSelRF'' package from  \cite{diaz2007genesrf} is then used in default mode, except for the fraction of variables to exclude from the prior forest at each iteration. Because there are so few important variables, it is set to 0.5. The ``Boruta'' package from \cite{kursa2010feature} is then used for the Boruta algorithm, with the maximum number of importance source runs set to 1000. Variables are chosen in three steps for VSURF. The algorithm is used to determine the features that survive the prediction stage. The number of trees in each forest is set to 500 while implementing it with the ``VSURF'' package (\cite{vsurf}). VSURF's default parallelization setting is false. We convert it to TRUE to speed up computations. RRF is implemented using ``RRF'' package (\cite{deng2012feature}), with $\gamma$ set to 0.5 to obtain regularization coefficients. Finally, the ``vita'' package \cite{vita}) is used to implement the PIMP and NTA methods. Both techniques would consider features with $p$-values less than 0.05 to be picked.

All the codes are run in RStudio Version 1.4.1717, equipped with R version 4.1.0, on a laptop with 8GB RAM and an 8-core AMD processor. Barring the specifications above, other arguments in all the packages are used in the default settings. 
All the functions execute on a single core, with the exception of SIVS and VSURF, which have parallelization enabled by the function itself.

\subsection{Evaluation metrics}
\label{subsubsec:metrics}

In the simulation study, we use ten different metrics to evaluate the performance of various methods. As we pointed out before, our objective is to assess the parsimony and correctness of the variable selection methods and their accuracy in estimation and prediction. 

In order to address the first aspect above, we find the number of selected variables in each iteration and calculate the average across all iterations. It helps us to identify if a method performs well in selecting the correct number of important covariates. However, the correct number of covariates does not necessarily imply the correct set of covariates. We compute the average proportions of important and unimportant variables selected to analyse this particular property. A good procedure is expected to attain a near-correct selected number of variables. Further, the percentage of important variables selected should be close to 100\%, while the same for the unimportant variables is expected to be close to zero. To define these quantities mathematically, let $\hat\beta_j$ and $\beta_j$ (for $1\leqslant j \leqslant m$) be the estimated and the true coefficients. We use $\hat{\bm\beta}$ and $\bm\beta$ to denote the corresponding vectors. Then, the aforementioned quantities are defined as
\begin{equation}
\label{eq:selected-imp-unimp}
\begin{split}
    \mathrm{Selected} &= \norm{\hat{\bm\beta}}_0, \\ 
    \mathrm{Imp\%} &= \frac{\#\left\{ 1 \leqslant j \leqslant m : \hat\beta_j \neq 0, \beta_j\neq 0 \right\}}{\#\left\{ 1 \leqslant j \leqslant m : \beta_j\neq 0 \right\}}, \\ 
    \mathrm{Unimp\%} &= \frac{\#\left\{ 1 \leqslant j \leqslant m : \hat\beta_j \neq 0, \beta_j = 0 \right\}}{\#\left\{ 1 \leqslant j \leqslant m : \beta_j = 0 \right\}}.
\end{split}
\end{equation}

Comparison of the estimation performance of the methods is done through two different measures, which are remarkably popular in statistical literature (\cite{chai2014root}). These are mean squared error (MSE) and mean absolute error (MAE). For both measures, we shall report the median values across all iterations. Because tree-based methods are not typically used for estimation purposes, we calculate these errors only for the other typologies. The formal definitions of these measures are provided below.
\begin{equation}
\label{eq: mse-mae}
\mathrm{MSE} = \frac{1}{n}  \sum_{j=1}^m(\beta_j - \hat{\beta_j})^2, \; \mathrm{MAE} = \frac{1}{n}  \sum_{j=1}^m \abs{\beta_j - \hat{\beta_j}}.
\end{equation}

For assessing the prediction accuracy of different methods, in each experiment, the data is divided into a training set and a test set, the latter being denoted as $S_{\mathrm{te}}$. We fit each method to the training set and use that to predict the probability distribution for the observations in $S_{\mathrm{te}}$. Below, $\hat\pi_i$ denotes the predicted probability of success for the $i^{th}$ event in the test set.

Now, to evaluate the prediction performance, we first compute the empirical accuracy as the percentage of overall correct predictions where success is predicted by $(\hat\pi_i>0.5)$. Next, we calculate the precision and the recall. The first one measures the accuracy of positives classified by the method, whereas recall measures the ability of a method to classify positives correctly. Both have their advantages in judging the prediction accuracy of a method. These measures lie between 0 and 1, and higher values are preferred. If TP, FP, and FN denote the true positives, the false positives and the false negatives respectively, then the above three quantities are defined as follows:
\begin{equation}
\label{eq: accuracy}
\mathrm{Accuracy} = 100\left(\mathrm{\frac{TP + TN}{TP + FP + TN + FN}}\right), \; \mathrm{Precision} = \mathrm{\frac{TP}{TP + FP}}, \; \mathrm{Recall} = \mathrm{\frac{TP}{TP + FN}}.
\end{equation}

Evidently, the above measures give an idea of the classification abilities of the method. However, they cannot provide an adequate idea of the predictive accuracy in the LRM setup where not only the predicted category, but also the predictive distribution is crucial (\cite{czado2009predictive}). To tackle this aspect, we use the Brier score (\cite{brier1950verification}) and show the closeness of the predicted probability distribution to the observed outcome. It is a proper scoring rule and is defined as 
\begin{equation}
   \mathrm{Brier} = \frac{1}{\abs{S_{\mathrm{te}}}}\sum_{i\in S_{\mathrm{te}}}(y_i - \hat\pi_i)^2.
\end{equation}

Finally, we compute the average time taken by each variable selection method in each iteration under different settings. Computational time is always a pivotal metric in such studies, for a time-consuming method can turn out to be an infeasible choice in high-dimensional cases.

\subsection{Simulation results}
\label{subsec: simulation results}

In \Cref{fig:varsel_diag} below, we first provide a concise outline of recommended methods based on their ability to select variables and predict outcomes in different dimensional setup. The methods are classified into smaller subgroups based on the similarities and dissimilarities in their behaviors, and we indicate whether each subgroup is suitable in terms of the aforementioned criteria for the three settings. For instance, one can see that the popular variable selection methods LASSO, ALASSO and ElasticNet are suitable for every setting in terms of their predictive abilities, but their selection accuracy are not great. Furthermore, in each subgroup, we include some related points that might help readers to choose the best method for their needs. Overall, based on the simulation results, we recommend that SIVS should be used in a low-dimensional setup whereas in both moderate and high-dimensional setups, MCP is arguably the most appropriate method. In case of the latter settings, the Best Subset approach can also be used if the sample size $n$ is large.

\begin{figure}[!ht]
\centering 
\includegraphics[width =  \textwidth]{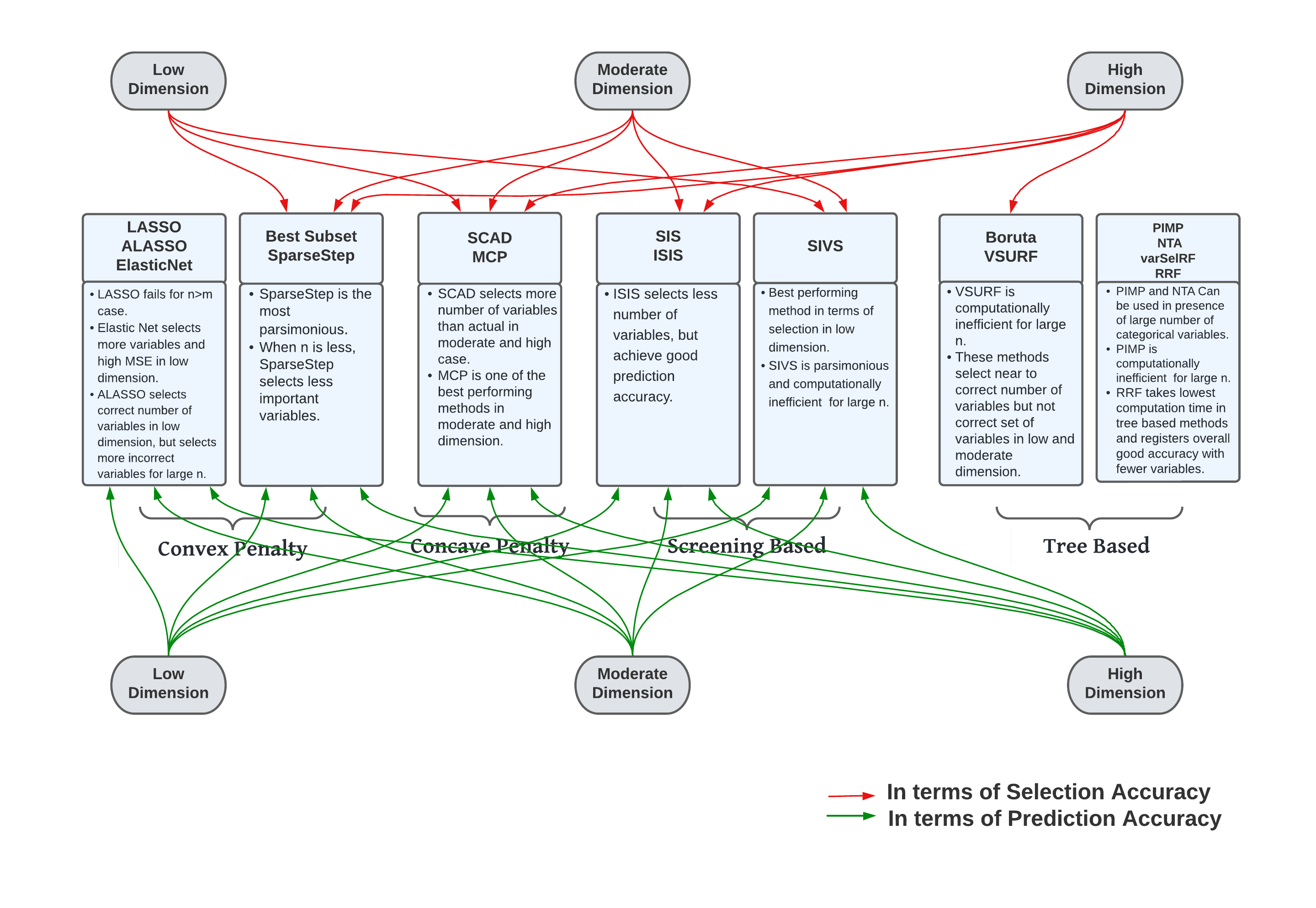}
\caption{Recommendations based on selection and prediction accuracy}
\label{fig:varsel_diag} 
\end{figure}



\begin{figure}[!ht]
\centering 
\includegraphics[width = 0.9\textwidth]{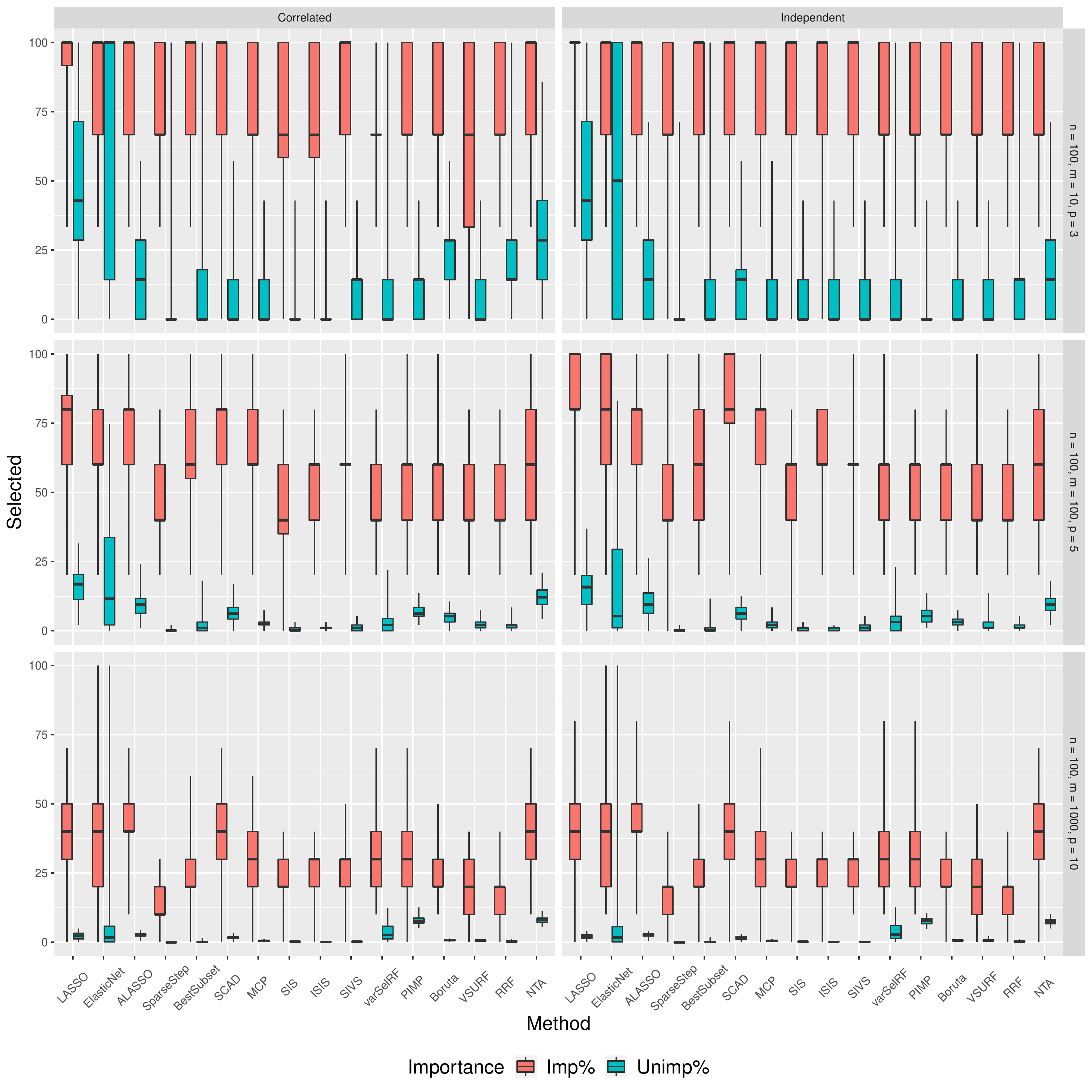}
\caption{Boxplots of the percentage of important and unimportant variables selected by each method across various iterations in case of $n = 100$. Correlated cases are shown in the left panel and independent cases are shown in the right panel for different dimensions.}
\label{fig:boxplot100} 
\end{figure}

\begin{figure}[!ht]
\centering 
\includegraphics[width = 0.9\textwidth]{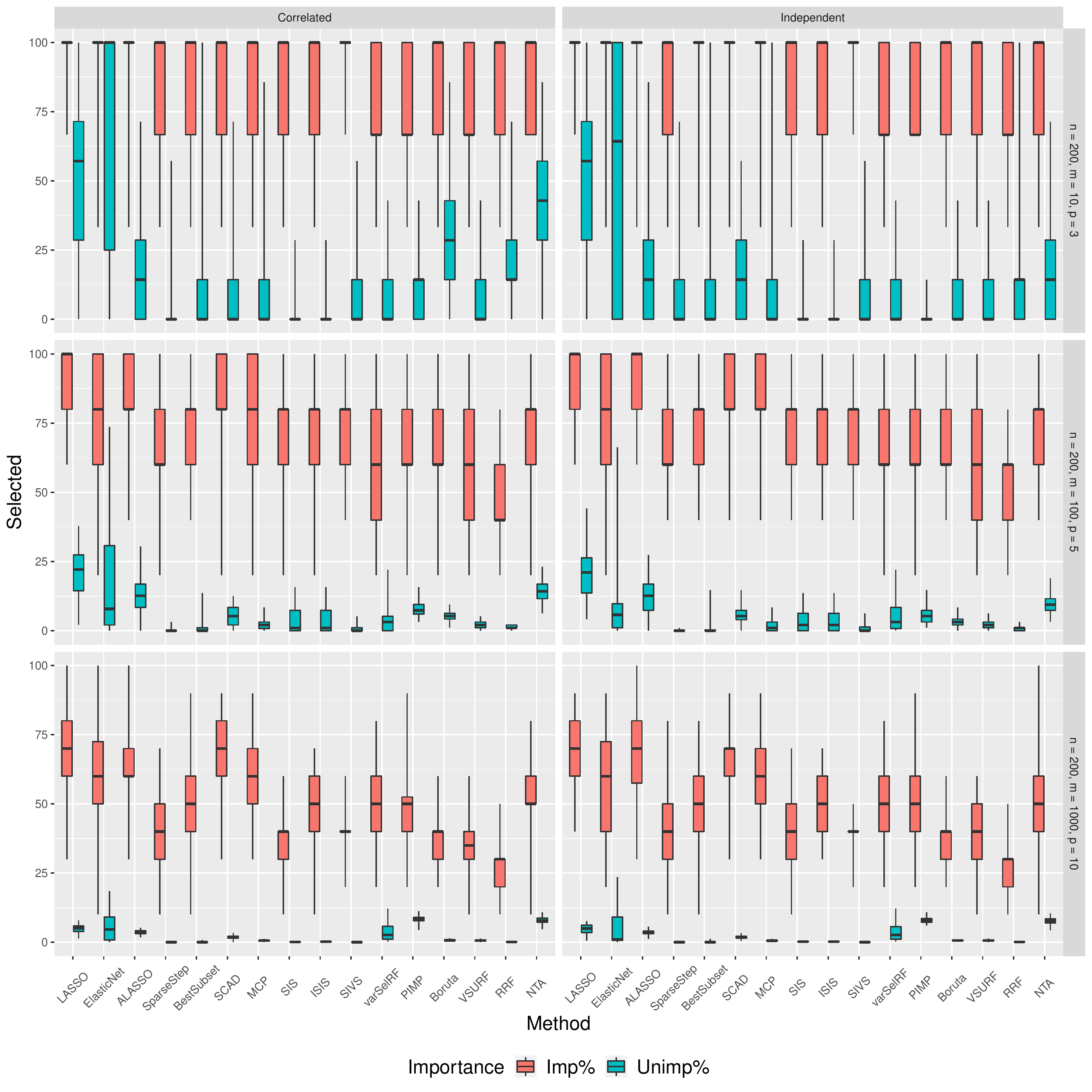}
\caption{Boxplots of the percentage of important and unimportant variables selected by each method across various iterations in case of $n = 200$. Correlated cases are shown in the left panel and independent cases are shown in the right panel for different dimensions.}
\label{fig:boxplot200} 
\end{figure}

\begin{figure}[!ht]
\centering 
\includegraphics[width = 0.9\textwidth]{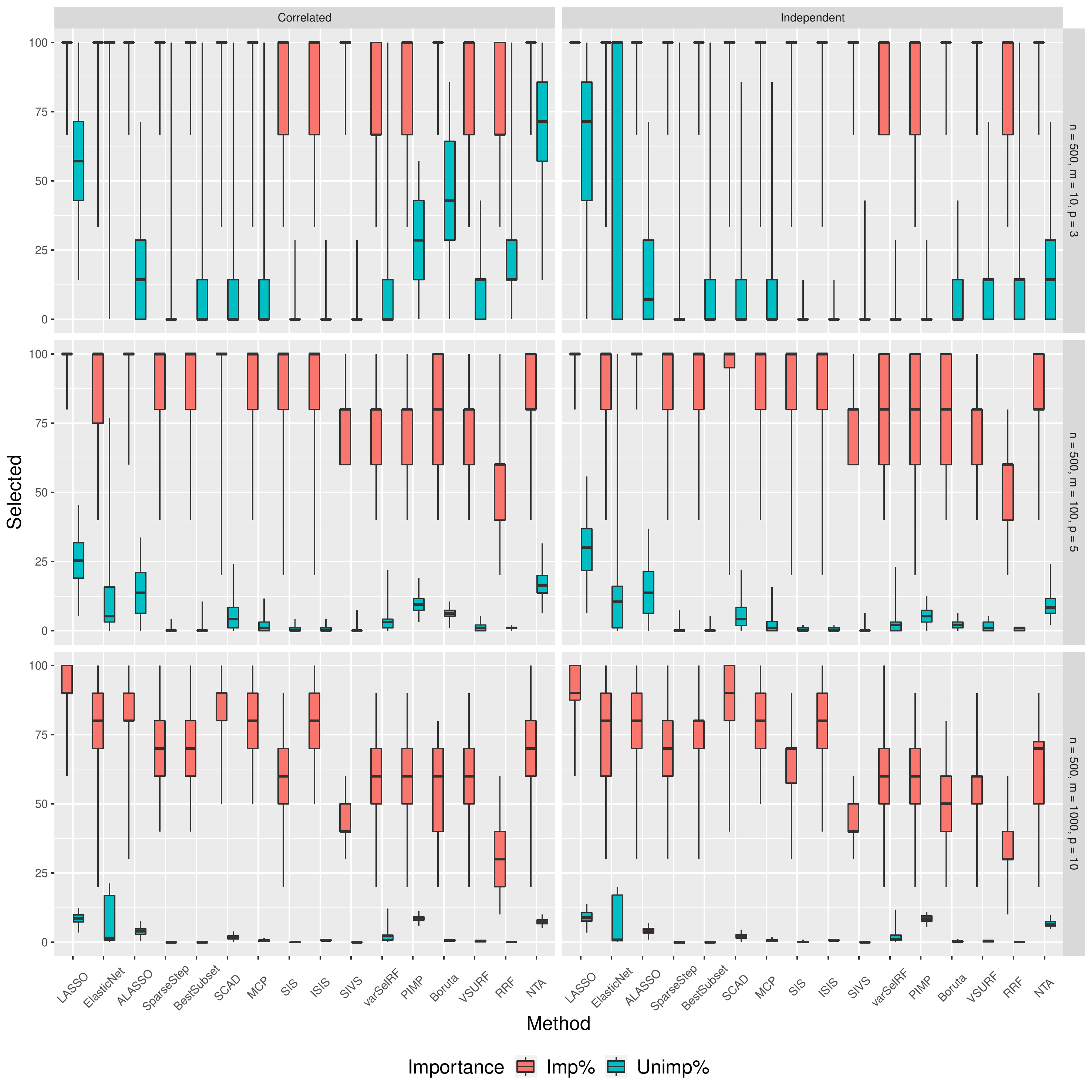}
\caption{Boxplots of the percentage of important and unimportant variables selected by each method across various iterations in case of $n = 500$. Correlated cases are shown in the left panel and independent cases are shown in the right panel for different dimensions.}
\label{fig:boxplot500} 
\end{figure}

Let us now go over the simulation results in more detail. We look at three important metrics (selected number of variables, MSE and prediction accuracy) for all the methods in all setups. These are presented in tables \ref{tab:simulation-summary-results1}, \ref{tab:simulation-summary-results2} and \ref{tab:simulation-summary-results3}, for different choices of $n$. For the sake of brevity, detailed results on all other metrics are deferred to the Appendix (\Cref{tab:setup11} to \Cref{tab:setup19}). We also look at the percentage of important and unimportant variables selected by each method in all the setups (figures \ref{fig:boxplot100}, \ref{fig:boxplot200} and \ref{fig:boxplot500}, shown in \Cref{sec:figures}). Because all of the setups are shown on the same scale in these figures, the values corresponding to Unimp\% in box plots drops to a very low number with the increment in $m$. These graphs show how the percentage fluctuates across different settings for a fixed $n$. To get the exact values of these percentages, please see the aforementioned tables in the Appendix.

\begin{table}[!ht]
\centering
\caption{Three important performance metrics for all the methods under different simulation setups in case of $n=100$.}
\label{tab:simulation-summary-results1}
\scalebox{0.85}{
\begin{tabular}{lccccccccc}
    \hline
  & \multicolumn{9}{c}{Independent case} \\
  Setup & \multicolumn{ 3 }{c}{ $m = 10, p =3$} & \multicolumn{ 3 }{c}{ $m = 100, p =5$} & \multicolumn{ 3 }{c}{ $m = 1000, p =10$}\\
\hline
 Method & Selected & MSE  & Accuracy & Selected & MSE  & Accuracy & Selected & MSE  & Accuracy \\ 
 \hline
 LASSO & 6.43 & 0.38 & 83.30 & 18.43 & 0.19 & 83.45 & 24.31 & 0.06 & 74.15 \\ 
  ElasticNet & 6.33 & 1.60 & 82.30 & 17.63 & 0.37 & 79.65 & 53.54 & 0.07 & 69.80 \\ 
  ALASSO & 3.91 & 0.21 & 83.90 & 13.14 & 0.11 & 83.25 & 30.54 & 0.05 & 72.75 \\ 
  SparseStep & 2.54 & 0.22 & 84.00 & 2.61 & 0.10 & 82.35 & 1.80 & 0.05 & 69.65 \\ 
  Best Subset & 3.67 & 0.24 & 84.15 & 3.85 & 0.11 & 82.85 & 4.93 & 0.06 & 71.70 \\ 
  \hline
  SCAD & 3.47 & 0.29 & 84.35 & 9.95 & 0.13 & 84.20 & 19.35 & 0.06 & 74.30 \\ 
  MCP & 3.09 & 0.30 & 84.65 & 5.78 & 0.14 & 84.00 & 8.01 & 0.06 & 74.95 \\ 
  \hline
  SIS & 2.71 & 0.38 & 83.60 & 3.35 & 0.12 & 82.35 & 3.81 & 0.05 & 70.70 \\ 
  ISIS & 2.71 & 0.38 & 83.60 & 3.85 & 0.07 & 82.60 & 3.91 & 0.05 & 71.90 \\ 
  SIVS & 3.08 & 0.80 & 84.57 & 4.06 & 0.20 & 82.55 & 4.19 & 0.06 & 72.10 \\ 
  \hline
  varSelRF & 2.67 &  & 80.75 & 6.45 &  & 76.50 & 42.42 &  & 63.85 \\ 
  PIMP & 2.45 &  & 80.89 & 7.89 &  & 75.75 & 79.52 &  & 61.95 \\ 
  Boruta & 2.86 &  & 81.70 & 5.68 &  & 75.51 & 8.46 &  & 64.15 \\ 
  VSURF & 2.77 &  & 81.12 & 4.37 &  & 76.77 & 8.55 &  & 63.55 \\ 
  RRF & 3.40 &  & 80.74 & 3.67 &  & 76.65 & 5.22 &  & 67.25 \\ 
  NTA & 3.43 &  & 80.37 & 12.36 &  & 74.60 & 77.85 &  & 62.55 \\
 \hline
   & \multicolumn{9}{c}{Correlated case} \\
  Setup & \multicolumn{ 3 }{c}{ $m = 10, p =3$} & \multicolumn{ 3 }{c}{ $m = 100, p =5$} & \multicolumn{ 3 }{c}{ $m = 1000, p =10$} \\
\hline
Method & Selected & MSE  & Accuracy & Selected & MSE  & Accuracy & Selected & MSE  & Accuracy \\ 
 \hline
 LASSO & 5.91 & 0.41 & 82.60 & 19.20 & 0.19 & 81.00 & 26.22 & 0.06 & 76.35 \\ 
  ElasticNet & 6.95 & 1.42 & 81.65 & 21.89 & 0.36 & 78.05 & 63.30 & 0.07 & 71.10 \\ 
  ALASSO & 3.75 & 0.30 & 83.75 & 12.45 & 0.12 & 81.60 & 29.70 & 0.05 & 73.80 \\ 
  SparseStep & 2.35 & 0.21 & 83.05 & 2.60 & 0.11 & 80.90 & 1.59 & 0.05 & 71.50 \\ 
  Best Subset & 3.51 & 0.34 & 83.10 & 5.40 & 0.13 & 81.50 & 5.24 & 0.06 & 73.10 \\ 
  \hline
  SCAD & 3.20 & 0.40 & 82.45 & 9.90 & 0.13 & 82.80 & 20.26 & 0.06 & 75.55 \\ 
  MCP & 2.78 & 0.35 & 82.80 & 5.89 & 0.12 & 82.65 & 8.06 & 0.06 & 75.85 \\ 
  \hline
  SIS & 2.45 & 0.52 & 82.05 & 2.61 & 0.17 & 77.30 & 3.55 & 0.05 & 73.10 \\ 
  ISIS & 2.45 & 0.52 & 82.05 & 3.81 & 0.14 & 81.40 & 3.97 & 0.04 & 73.65 \\ 
  SIVS & 2.95 & 0.86 & 83.92 & 3.93 & 0.20 & 82.15 & 4.11 & 0.06 & 75.91 \\ 
  \hline
  varSelRF & 2.64 &  & 77.65 & 5.44 &  & 74.80 & 35.16 &  & 67.95 \\ 
  PIMP & 2.85 &  & 77.39 & 8.89 &  & 73.70 & 80.73 &  & 63.55 \\ 
  Boruta & 3.91 &  & 79.11 & 7.04 &  & 73.85 & 9.92 &  & 68.20 \\ 
  VSURF & 2.75 &  & 78.15 & 4.10 &  & 74.85 & 7.94 &  & 66.95 \\ 
  RRF & 3.74 &  & 78.76 & 3.96 &  & 74.35 & 5.20 &  & 67.25 \\ 
  NTA & 4.79 &  & 78.58 & 14.51 &  & 74.20 & 82.59 &  & 63.75 \\
  \hline
\end{tabular}
}
\end{table}

\begin{table}[!ht]
\centering
\caption{Three important performance metrics for all the methods under different simulation setups in case of $n=200$.}
\label{tab:simulation-summary-results2}
\scalebox{0.85}{
\begin{tabular}{lccccccccc}
    \hline
  & \multicolumn{9}{c}{Independent case} \\
  Setup & \multicolumn{ 3 }{c}{ $m = 10, p =3$} & \multicolumn{ 3 }{c}{ $m = 100, p =5$} & \multicolumn{ 3 }{c}{ $m = 1000, p =10$}\\
\hline
 Method & Selected & MSE  & Accuracy & Selected & MSE  & Accuracy & Selected & MSE  & Accuracy \\ 
 \hline
 LASSO & 6.55 & 0.20 & 85.28 & 24.61 & 0.10 & 85.60 & 53.70 & 0.06 & 82.70 \\ 
  ElasticNet & 6.76 & 0.78 & 84.53 & 15.90 & 0.30 & 82.80 & 47.25 & 0.05 & 80.28 \\ 
  ALASSO & 3.84 & 0.14 & 85.22 & 16.26 & 0.04 & 85.17 & 42.14 & 0.04 & 82.08 \\ 
  SparseStep & 3.14 & 0.10 & 85.50 & 3.45 & 0.03 & 85.92 & 4.36 & 0.03 & 84.45 \\ 
  Best Subset & 3.83 & 0.14 & 85.05 & 4.19 & 0.05 & 85.53 & 5.82 & 0.04 & 84.97 \\ 
  \hline
  SCAD & 3.72 & 0.11 & 85.15 & 9.58 & 0.05 & 86.10 & 24.68 & 0.03 & 86.70 \\ 
  MCP & 3.37 & 0.11 & 85.40 & 6.19 & 0.04 & 86.47 & 11.88 & 0.03 & 86.85 \\ 
  \hline
  SIS & 2.78 & 0.15 & 85.42 & 7.36 & 0.23 & 84.17 & 6.08 & 0.04 & 81.03 \\ 
  ISIS & 2.78 & 0.15 & 85.42 & 7.36 & 0.23 & 84.17 & 6.97 & 0.02 & 84.15 \\ 
  SIVS & 3.06 & 0.62 & 85.56 & 4.61 & 0.12 & 85.55 & 4.34 & 0.07 & 83.28 \\ 
  \hline
  varSelRF & 2.90 &  & 82.53 & 8.00 &  & 80.80 & 37.54 &  & 73.58 \\ 
  PIMP & 2.56 &  & 82.67 & 8.30 &  & 80.60 & 83.43 &  & 71.78 \\ 
  Boruta & 3.04 &  & 82.47 & 6.14 &  & 81.83 & 9.77 &  & 75.33 \\ 
  VSURF & 3.05 &  & 82.55 & 4.89 &  & 81.20 & 9.73 &  & 76.65 \\ 
  RRF & 3.35 &  & 82.27 & 3.30 &  & 80.18 & 3.90 &  & 75.95 \\ 
  NTA & 3.59 &  & 82.83 & 12.69 &  & 79.75 & 81.06 &  & 71.42 \\ 
 \hline
   & \multicolumn{9}{c}{Correlated case} \\
  Setup & \multicolumn{ 3 }{c}{ $m = 10, p =3$} & \multicolumn{ 3 }{c}{ $m = 100, p =5$} & \multicolumn{ 3 }{c}{ $m = 1000, p =10$} \\
\hline
Method & Selected & MSE  & Accuracy & Selected & MSE  & Accuracy & Selected & MSE  & Accuracy \\ 
 \hline
LASSO & 6.51 & 0.20 & 84.78 & 24.31 & 0.11 & 85.62 & 54.32 & 0.06 & 82.50 \\ 
  ElasticNet & 7.74 & 0.38 & 83.62 & 19.34 & 0.31 & 82.10 & 57.29 & 0.05 & 78.83 \\ 
  ALASSO & 3.93 & 0.13 & 84.62 & 16.14 & 0.06 & 85.85 & 41.28 & 0.04 & 82.33 \\ 
  SparseStep & 2.90 & 0.09 & 84.10 & 3.46 & 0.03 & 86.78 & 4.38 & 0.03 & 83.92 \\ 
  Best Subset & 3.57 & 0.11 & 84.10 & 4.26 & 0.04 & 86.50 & 5.74 & 0.04 & 84.38 \\
  \hline
  SCAD & 3.38 & 0.14 & 84.22 & 9.40 & 0.05 & 87.12 & 24.58 & 0.03 & 85.65 \\ 
  MCP & 3.14 & 0.12 & 84.50 & 6.24 & 0.04 & 86.97 & 11.76 & 0.04 & 85.40 \\ 
  \hline
  SIS & 2.68 & 0.15 & 84.05 & 6.94 & 0.18 & 85.60 & 4.99 & 0.05 & 80.92 \\ 
  ISIS & 2.68 & 0.15 & 84.05 & 6.94 & 0.18 & 85.60 & 6.94 & 0.02 & 83.80 \\ 
  SIVS & 2.92 & 0.61 & 84.62 & 4.38 & 0.12 & 87.10 & 4.24 & 0.06 & 83.20 \\
  \hline
  varSelRF & 2.74 &  & 81.65 & 6.82 &  & 81.83 & 37.50 &  & 74.40 \\ 
  PIMP & 3.33 &  & 81.28 & 10.40 &  & 81.00 & 86.97 &  & 71.38 \\ 
  Boruta & 4.64 &  & 81.31 & 8.42 &  & 81.45 & 11.06 &  & 75.03 \\ 
  VSURF & 2.95 &  & 81.61 & 4.84 &  & 81.34 & 9.69 &  & 74.45 \\ 
  RRF & 3.78 &  & 81.26 & 3.57 &  & 80.90 & 4.24 &  & 74.47 \\ 
  NTA & 5.67 &  & 80.99 & 16.95 &  & 79.95 & 84.07 &  & 72.08 \\   
  \hline
\end{tabular}
}
\end{table}

\begin{table}[!ht]
\centering
\caption{Three important performance metrics for all the methods under different simulation setups in case of $n=500$.}
\label{tab:simulation-summary-results3}
\scalebox{0.85}{
\begin{tabular}{lccccccccc}
    \hline
  & \multicolumn{9}{c}{Independent case} \\
  Setup & \multicolumn{ 3 }{c}{ $m = 10, p =3$} & \multicolumn{ 3 }{c}{ $m = 100, p =5$} & \multicolumn{ 3 }{c}{ $m = 1000, p =10$}\\
\hline
 Method & Selected & MSE  & Accuracy & Selected & MSE  & Accuracy & Selected & MSE  & Accuracy \\ 
 \hline
 LASSO & 7.35 & 0.12 & 84.32 & 32.40 & 0.08 & 87.23 & 98.36 & 0.03 & 88.40 \\ 
  ElasticNet & 6.78 & 0.30 & 83.91 & 14.90 & 0.32 & 86.11 & 79.46 & 0.04 & 86.71 \\ 
  ALASSO & 3.97 & 0.06 & 84.59 & 18.55 & 0.02 & 87.23 & 48.74 & 0.02 & 89.06 \\ 
  SparseStep & 3.54 & 0.04 & 84.32 & 4.67 & 0.01 & 87.98 & 7.31 & 0.01 & 90.84 \\ 
  Best Subset & 3.61 & 0.06 & 84.28 & 4.87 & 0.02 & 87.90 & 8.16 & 0.01 & 90.93 \\ 
  \hline
  SCAD & 3.59 & 0.06 & 83.98 & 10.11 & 0.01 & 88.01 & 29.38 & 0.01 & 91.00 \\ 
  MCP & 3.33 & 0.05 & 84.32 & 7.01 & 0.01 & 88.19 & 14.95 & 0.01 & 91.23 \\
  \hline
  SIS & 2.77 & 0.06 & 84.23 & 4.77 & 0.03 & 87.34 & 8.08 & 0.02 & 88.85 \\ 
  ISIS & 2.77 & 0.06 & 84.23 & 4.77 & 0.03 & 87.34 & 14.99 & 0.04 & 88.89 \\ 
  SIVS & 3.05 & 0.36 & 84.54 & 4.15 & 0.13 & 86.39 & 4.35 & 0.05 & 84.73 \\ 
  \hline
  varSelRF & 3.09 &  & 81.56 & 7.30 &  & 84.07 & 28.59 &  & 81.68 \\ 
  PIMP & 2.80 &  & 81.43 & 8.83 &  & 84.30 & 89.88 &  & 78.68 \\ 
  Boruta & 3.35 &  & 81.84 & 5.92 &  & 83.99 & 8.43 &  & 82.44 \\ 
  VSURF & 3.52 &  & 81.73 & 5.33 &  & 84.20 & 9.58 &  & 82.56 \\ 
  RRF & 3.56 &  & 81.55 & 3.47 &  & 82.12 & 4.12 &  & 78.74 \\ 
  NTA & 4.08 &  & 82.05 & 12.99 &  & 83.95 & 73.66 &  & 79.18 \\ 
 \hline
   & \multicolumn{9}{c}{Correlated case} \\
  Setup & \multicolumn{ 3 }{c}{ $m = 10, p =3$} & \multicolumn{ 3 }{c}{ $m = 100, p =5$} & \multicolumn{ 3 }{c}{ $m = 1000, p =10$} \\
\hline
Method & Selected & MSE  & Accuracy & Selected & MSE  & Accuracy & Selected & MSE  & Accuracy \\ 
 \hline
 LASSO & 7.11 & 0.13 & 84.85 & 29.45 & 0.07 & 87.97 & 92.09 & 0.04 & 88.28 \\ 
  ElasticNet & 9.00 & 0.08 & 84.39 & 17.63 & 0.33 & 85.12 & 92.87 & 0.04 & 85.91 \\ 
  ALASSO & 4.24 & 0.08 & 84.90 & 17.84 & 0.02 & 87.99 & 47.06 & 0.02 & 89.13 \\ 
  SparseStep & 3.45 & 0.05 & 84.87 & 4.61 & 0.01 & 87.99 & 7.36 & 0.01 & 90.38 \\ 
  Best Subset & 3.53 & 0.07 & 84.98 & 4.90 & 0.02 & 87.99 & 7.79 & 0.01 & 90.73 \\ 
  \hline
  SCAD & 3.60 & 0.08 & 84.35 & 9.46 & 0.02 & 87.90 & 26.26 & 0.01 & 90.81 \\ 
  MCP & 3.40 & 0.05 & 84.81 & 6.44 & 0.01 & 87.92 & 13.80 & 0.01 & 90.80 \\
  \hline
  SIS & 2.83 & 0.07 & 84.62 & 4.80 & 0.04 & 87.65 & 6.94 & 0.02 & 88.91 \\ 
  ISIS & 2.83 & 0.07 & 84.62 & 4.80 & 0.04 & 87.65 & 14.64 & 0.07 & 88.85 \\ 
  SIVS & 3.02 & 0.42 & 84.76 & 4.20 & 0.12 & 86.64 & 4.36 & 0.05 & 85.21 \\ 
  \hline
  varSelRF & 3.03 &  & 82.27 & 7.30 &  & 83.80 & 30.81 &  & 81.46 \\ 
  PIMP & 4.46 &  & 82.47 & 12.92 &  & 83.65 & 91.29 &  & 79.20 \\ 
  Boruta & 6.12 &  & 82.68 & 10.13 &  & 84.15 & 12.05 &  & 82.12 \\ 
  VSURF & 3.38 &  & 82.94 & 4.88 &  & 84.19 & 9.54 &  & 82.46 \\ 
  RRF & 3.90 &  & 82.06 & 3.68 &  & 81.83 & 4.26 &  & 78.76 \\ 
  NTA & 7.27 &  & 81.82 & 20.11 &  & 82.83 & 80.21 &  & 79.55 \\  
  \hline
\end{tabular}
}
\end{table}

Across all eighteen simulation configurations, we see that the tree-based approaches have the lowest efficacy in selecting appropriate variables. In case of a fixed value of $n$, the selection performance for all methods in each setup declines as the number of variables grows. Except for ElasticNet, PIMP, Boruta, and NTA, the performances of the methods in terms of the number of selected variables is not significantly different from independent to correlated setup. In the correlated scenario, those four methods often select more variables than in the independent case. However, the values of Imp\% and Unimp\% from figures \ref{fig:boxplot100} to \ref{fig:boxplot500} show some variability. ElasticNet, PIMP, Boruta, and NTA have more fluctuation since they select a larger number of variables in correlated cases. One probable explanation is that in the situation of correlated variables, if two variables are significantly associated but only one is significant, a few approaches cannot make the proper choice. Alternatively, if multiple variables are substantially connected and more than one variable is relevant, some methods fail to make the correct choice and select only one variable.

Except for LASSO in the independent case and SCAD in the correlated scenario, for $n = 100$, all approaches' prediction accuracy declines as the number of variables increases. However the behavior changes in case of different values of $n$. For a sample size of 200 under the independent setup, as the number of variables increases, the prediction accuracy of convex penalty methods SCAD and MCP increases whereas it drops for ALASSO, ElasticNet, and all tree-based and screening-based methods. In the correlated setup though, all methods record almost similar accuracy in low and moderate dimensions but we see a decline in the high-dimensional case. 

Finally, under the setup of $n = 500$, all penalty-based and screening-based methods except SIVS improve prediction accuracy as the number of variables increases in all setups. In the correlated scenario, the prediction accuracy of RRF decreases as the number of variables increases. In other cases not covered above, accuracy increases as the number of variables increases from 10 to 100 and decreases as the number of variables increases from 100 to 1000. 

One intriguing finding is that in a low-dimensional configuration, the MSE values for estimating the coefficients for ElasticNet are quite high. As a result, even though the selection patterns and accuracy metrics are good, we might argue that the estimated coefficients in the low-dimensional setting from ElasticNet should not be trusted.

For a fixed value of $m$, one can observe that in most situations, the number of variables selected by a method rises as the value of $n$ increases. In the low-dimension case, we find that only LASSO and ElasticNet pick a larger number of variables. When we increase the value of $m$ to 100, we notice that ALASSO, PIMP, and NTA pick a greater number of variables, with a faster increment rate than LASSO and ElasticNet. In fact, the number of selected variables by ElasticNet decreases as the number of observations increases when $m=100$. Eventually in the high dimensional case, the increment rates in picking variables for PIMP and NTA becomes exceptionally high. According to our findings, only MCP, Boruta, and VSURF select close to the correct number of variables. On the other hand, LASSO, ALASSO, ElasticNet, SCAD, VarSelRF, PIMP and NTA select a greater number of variables in general, whereas SparseStep, Best Subset, and RRF select only a few in all iterations. In this regard, screening-based approaches are most conservative. 

\begin{figure}[!ht]
\centering 
\includegraphics[width = 0.9\textwidth]{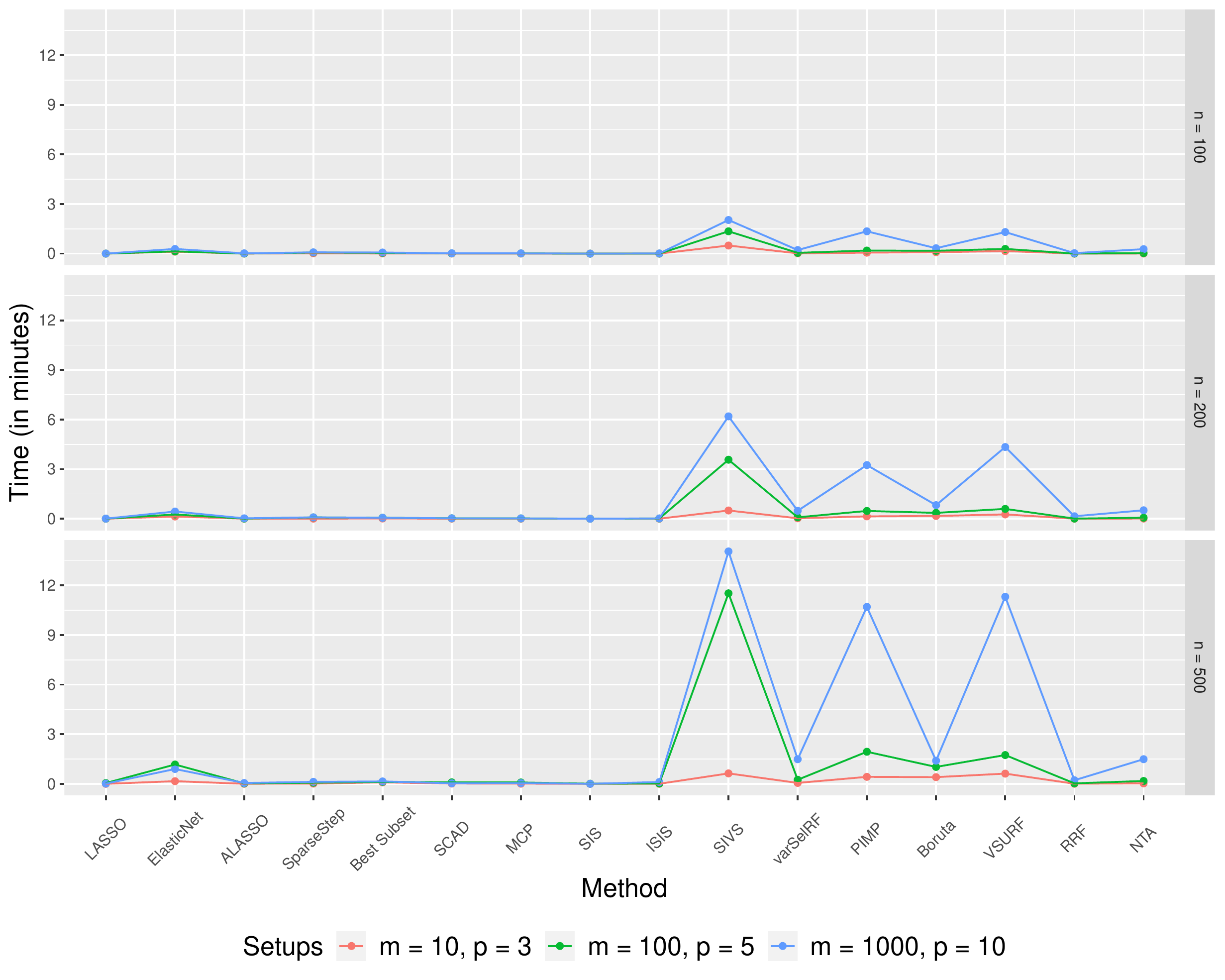}
\caption{Average computation time for the sixteen methods corresponding to the different setups in the simulation study. The computation time is shown in the panels above for $n = 100$, $n = 200$, and $n = 500$, respectively.}
\label{fig:time} 
\end{figure}

As a last piece in this section, we investigate the time complexity of the sixteen methods under different setups. Since the computational challenges are not dependent on the correlation structures of the regressors, we calculate the average time for all the methods for different choices of $n,m,p$, taking the correlated and the independent case together. These values are displayed in \Cref{fig:time} in \Cref{sec:figures}, and one can notice that the time consumption increases as the number of features increases. Except for ElasticNet, all penalty-based approaches take extremely little time in all situations. ElasticNet's computational burden is higher because it selects more variables most of the time. SIVS, when compared to other screening-based approaches, requires a significant amount of time. It is worth noting that as the value of $n$ rises, so does the computation time for SIVS. We should also mention that VSURF and PIMP are the most expensive among the tree-based approaches, although VarSelRF and NTA also require a long time. RRF is the quickest tree-based approach, and its complexity does not vary much depending on the situation. When the number of variables is 100, we see that SIVS takes a longer time than other approaches. In contrast, when $m =1000$, SIVS takes a long time at first, but when $n$ increases, the rate of increase in computing time for VSURF and PIMP is relatively higher.

\section{An application to a real dataset}
\label{sec:real-analysis}

In this section, we consider a high dimensional classification dataset that uses the gene expression data of prostate cancer patients. Originally from the study of \cite{singh2002gene}, this dataset is freely available in the R package sda (\cite{Miika2021sda}) and has been widely studied by many researchers (see e.g.\ \cite{efron2009empirical}, \cite{genuer2010variable}). The dataset contains relevant information of 6033 genes for 102 subjects -- 52 prostate cancer patients and 50 healthy men. It falls within the regime of the LRM, as the response variable can be taken as whether a subject is a cancer patient or not, and the objective is to identify which genes are responsible for prostate cancer. 

For this dataset, we analyze and compare the performances of all of the variable selection methods discussed and explored in the previous sections. To begin with, using the complete data, we investigate the number of selected variables, running time and the accuracy in fitting the data for the sixteen methods. These results are detailed in \Cref{tab:performance}. Note that, because the true set of important variables are unknown, we can only present the selected number of variables, Brier score for the fitted probability distributions and the time taken by each method. To further understand how well the models fit the data, we calculate and report the area under the curve (AUC) as well.

\begin{table}[!htb]
\centering
\caption {Comparison of the sixteen methods in the real data ($n=102,m= 6033$).}
\label{tab:performance}
\begin{tabular}{lcccc}
  \hline
 Method & Selected & AUC & Brier & Time \\ 
  \hline
LASSO & 75 & 1 &  0.000 & 0.019  \\ 
ElasticNet & 875 &1 & 0.000& 1.058  \\ 
  ALASSO & 44 & 1 & 0.000& 0.118  \\ 
  SparseStep& 1 & 0.764 &0.185 & 0.450 \\ 
  Best Subset & 19 & 1 & 0.005 & 0.161  \\
  \hline
  SCAD & 50 &1 &0.018 & 0.102  \\ 
  MCP & 15 &0.999 & 0.057& 0.066  \\ 
  \hline
  SIS & 5 & 0.935& 0.092 & 0.006 \\ 
  ISIS & 5 & 0.956& 0.077& 0.120 \\ 
  SIVS & 5 & 0.945&0.088 & 4.731  \\ 
  \hline
  VarSelRF & 23 & 1 & 0.001 & 1.473 \\
  PIMP & 550 & 1& 0.003& 11.113  \\ 
  Boruta & 195 & 1&0.002 & 4.209  \\ 
  VSURF & 13 & 1& 0.002& 179.289 \\
  RRF & 6 & 1 & 0.002& 0.291  \\
  NTA & 1325 & 1 & 0.004 & 4.907  \\ 
   \hline
\end{tabular}
\end{table}

Among the norm-based penalty methods, we notice that except for SparseStep, all other techniques register great accuracy in classifying the patients correctly. However, the number of variables selected varies substantially among all the methods. While LASSO and ALASSO select a moderately high number of variables, ElasticNet, potentially due to the properties of ridge regularization, selects as many as 875 variables and renders an immensely complex model. On the contrary, SparseStep selects only one gene, and it is still found to have the AUC value of 0.764. The Best Subset method, in the sense of parsimony and accuracy, tends to be the most appropriate approach as it selects 19 variables and records an AUC value of 1 and a fitted Brier score of 0.005. 

Among the concave penalty-based methods, MCP is more parsimonious compared to SCAD, and both have near to one AUC score. We also find that all screening-based methods select only five genes and record good AUC and Brier scores, reflecting their usefulness in high dimensional data. Meanwhile, all the tree-based methods are found to fit the data very well. So far as the feature selection goes, except for RRF, VSURF and VarSelRF, other three techniques pick many genes to be important. They take considerable time to run as well. VSURF in particular is exceedingly time-consuming, which is inline with the discussion of time complexity in \Cref{subsec: simulation results}. Overall, one can consider VarSelRF as the most acceptable approach since it selects 23 variables and records an AUC of 1 and a fitted Brier score of 0.001, the lowest of all tree-based methods.

Next, to further demonstrate the effectiveness of these methods, we focus on the consistency in variable selection and estimation, as well as prediction performance. Here, we randomly select 80\% of the dataset as a training set and the remaining 20\% as a test set. This process is repeated for many iterations and the results are recorded. Due to the excessive computational burden, PIMP and VSURF are not considered here. For the other fourteen methods, in an attempt to evaluate how consistently a method determines the set of important variables across different iterations, we define a measure of similarity following the Sokal-Michener index (\cite{valsecchia2020similarity}).  Let us use $\hat\beta_{ik}$ to denote the estimated value of the $i^{th}$ coefficient in the $k^{th}$ iteration. Then the Sokal-Michener similarity index, for the $k^{th}$ and the $l^{th}$ iteration corresponding to a method, is computed as
\begin{equation}
\label{eqn: SM}
    \mathrm{SM}_{kl} =  \frac{\#\left\{ 1 \leqslant i \leqslant m : \left(\hat\beta_{ik} \neq 0 \; \text{and} \; \hat\beta_{il}\neq 0\right) \; \text{or} \; \left(\hat\beta_{ik} = 0 \; \text{and} \; \hat\beta_{il} = 0\right) \right\}}{m}.
\end{equation}

Using the above, if $r$ is the total number of iterations, then the global similarity index for a method is given as
\begin{equation}
\label{eqn: global-SM}
    \mathrm{SM} = \frac{1}{\binom{r}{2}}\sum_{k=1}^{r-1}\sum_{l=k+1}^{r} \mathrm{SM}_{kl}.
\end{equation}

Furthermore, to estimate how consistently a method estimates the coefficients of the variables across different iterations, we define the following consistency index (CI)
\begin{equation}
    \label{eqn:consistency}
    \mathrm{CI} = \frac{1}{m}\sum_{i = 1}^{m}\frac{1}{(r-1)}\sum_{j=1}^{r}\left(\hat\beta_{ij} - \frac{1}{r}\sum_{j=1}^r \hat\beta_{ij} \right)^2.
\end{equation}

The above measure signifies the average variation in the estimated coefficients across all iterations. Observe that the consistency index is not defined for tree-based methods as coefficient values are not estimated in those cases. Hence it is calculated for the other ten methods. A lower value of this index is desired to obtain consistency in estimating the coefficients of the variables.

\begin{table}[!htb]
\centering
\caption{Comparison of similarity and consistency in variable selection, estimation, prediction and computational complexity for all the methods (except VSURF and PIMP) across randomly generated train-test splits of the real data.}
\label{tab:measures}
\begin{tabular}{lccccc}
  \hline
 Method& Selected & Similarity & Consistency (in $10^{-4}$) & Time & Brier \\ 
  \hline
LASSO        & $51.15 \pm 6.59$ & 0.989 & 3.917 & 0.047 & 0.155 \\ 
  ALASSO     & $41.90 \pm 4.02$ & 0.992 & 8.045 & 0.234 & 0.152 \\ 
  SparseStep & $1.00 \pm 0.65$ & 0.999 & 3.219 & 0.246 & 0.244 \\ 
  ElasticNet & $630.95 \pm 306.18$ & 0.892 & 1.902 & 1.602 & 0.069 \\ 
  BestSubset & $18.30 \pm 1.45$ & 0.996 & 2.498 & 0.161 & 0.153 \\ 
  \hline
  SCAD       & $38.80 \pm 4.73$ & 0.992 & 0.079 & 0.058 & 0.153 \\ 
  MCP        & $11.55 \pm 2.31$ & 0.997 & 0.190 & 0.052 & 0.203 \\ 
  \hline
  SIS        & $4.00 \pm 0.32$ & 0.999 & 8.257 & 0.008 & 0.236 \\ 
  ISIS       & $4.10 \pm 0.45$ & 0.999 & 27.647 & 0.141 & 0.261 \\ 
  SIVS       & $4.25 \pm 0.55$ & 0.999 & 2.696 & 2.923 & 0.246 \\ 
  \hline 
  VarSelRF   & $17.30 \pm 10.59$ & 0.997 &  & 1.097 & 0.033 \\
  Boruta     & $161.55 \pm 6.89$ & 0.986 &  & 2.338 & 0.028 \\
  RRF        & $5.25 \pm 1.16$ & 0.999 &  & 0.147 & 0.064 \\ 
  NTA        & $1121.60 \pm 26.87$ & 0.810 &  & 1.642 & 0.042 \\ 
   \hline
\end{tabular}
\end{table}

\Cref{tab:measures} demonstrates that the behavior of all methods except Boruta, in terms of the number of variables selected, are pretty similar to the simulation study conducted for the high-dimensional data. SparseStep is identified to be the sparsest method. It typically selects only one gene in this real application. Best Subset, MCP, and VarSelRF pick a few number of regressors. Furthermore, we observe that the first two approaches select variables with low variability, while VarSelRF records greater variability in this regard. LASSO, ALASSO, and SCAD on average select moderate number of variables; whereas ElasticNet and NTA select a very high number of covariates. It is however interesting that despite selecting more than 1000 variables on an average, NTA is very consistent, i.e.\ there is minimal variation in the number of the selected genes by this method. Contrary to that, ElasticNet shows high variability in variable selection, although the estimated coefficients do not differ much. Finally, akin to the simulation study, screening-based methods tend to select only a handful of variables. It clearly depicts their tightness in picking only an appropriate and small set of variables in such a high dimensional data. 

The above observations are reaffirmed from the similarity index column of \Cref{tab:measures}. Methods with a high number of selected variables are generally not uniform in selecting the same set of genes across the iterations, while the screening-based methods are most homogeneous in selecting the important covariates. A conflicting point is noted from the consistency index which indicates that the estimated coefficients in screening-based typology are not homogeneous and that the concave penalty-based methods are the most dependable in this front. In fact, ISIS appears to be the most inconsistent method among all the non-tree-based methods in estimating the coefficients of the variables. 

The behaviors of the procedures of the last type are discovered to be dramatically different. RRF picks a small number of features but has a larger level of variability in variable selection than screening-based approaches. It is also one of the quickest algorithms. VarSelRF selects a modest amount of variables, but does so with greater variability than the Best subset and MCP techniques. Boruta and NTA, on the other hand, choose a lot of features in general and are more time-consuming approaches.

As a final piece of this section, we compare the prediction performance of the fourteen methods. The last column of \Cref{tab:measures} displays the average Brier scores across different iterations. We find that the screening-based methods are in general the least effective in predicting the true classes. In contrast, the tree-based methods and the ElasticNet are the best to classify the category of a patient in the out-of-sample data. Their Brier scores are, in truth, about four times better than the screening-based methods and more than two times better than other penalty-based methods. ElasticNet, possibly due to the selection of a bigger set of relevant genes, registers higher prediction accuracy than other penalty or screening-based methods. Among the tree-based methods, VarSelRF appears to be the most efficient choice. It generally selects sufficient number of genes to identify the category of the patients and registers high Brier score. In contrast, albeit Boruta marginally improves the prediction accuracy, it typically chooses almost ten times the number of genes that VarSelRF does. RRF, meanwhile, picks about one-third number of variables as compared to VarSelRF, but the error is nearly twice that of the latter. 

Based on the aforementioned findings, we make the following conjecture. In a medical study similar to above, if it is critical to identify all essential genes that may be linked to a disease, Boruta and VarSelRF may be the best approaches. The former selects a high number of variables and registers the best prediction performance, whereas the latter selects moderate number of variable without a significant drop in the accuracy. RRF is extremely restrictive in terms of variable selection and may not be suitable in similar applications. Amongst the penalty or screening based methods, only ElasticNet can be an effective method, and it usually chooses an exhaustive set of features that would provide good predictive performance.

\section{Concluding Remarks}
\label{sec:conclusion}

In this study, we have considered four types of classical variable selection methods in the regression model for the binary data, and assessed their efficacy in eighteen different simulation setups related to different dimensional cases of correlated and uncorrelated covariates. The accuracy of the methods deviate from one setting to another. For instance, in the low dimensional data where the number of observations is much higher than the number of predictors, SIVS, Best Subset, and MCP can be used. In these cases, LASSO and ElasticNet methods should be avoided as they select higher number of unimportant variables. In a moderate dimensional setting where the number of predictors is comparable to the number of observations, Best Subset and MCP still continue to perform well. SCAD can also be used in these cases as they are fast and accurate in out-of-sample predictions, albeit the former tends to select slightly greater number of covariates than the truth.

The results for the high-dimensional setting is more intriguing. We notice that the screening-based methods lean towards selecting very less number of variables. Thus, if the purpose is to choose a sparse yet good model, then these methods can be very useful. They provide acceptable prediction accuracy as well. Otherwise, concave penalty-based methods like SCAD and MCP can be good choices as they select appropriate number of total and important variables and record the lowest Brier scores in high dimensional setup. Based on the real application, we can also ascertain that Boruta and varSelRF are excellent choices in gene expression studies as they select appropriate number of important genes and registers the best prediction accuracy. On a related note, practitioners are often interested in detecting an exhaustive list of covariates which may be correlated with the binary response variable. In such research problems, ElasticNet or NTA, which tend to select a higher number of variables, can be more suitable algorithms for such requirements. Since LASSO can only select up to $\min\{n,m\}$ number of variables, it is not recommended in these problems, despite its predictive accuracy being at par with other penalty-based methods. We also note that SIVS, VSURF and PIMP take a lot of time to function even with appropriate parallelization, and thus they should be avoided while working with very high dimensional data with large $n$.

We want to conclude this paper with a succinct account of some future scopes of our work. Throughout the simulation study in this article, the number of important variables is taken to be substantially smaller than the number of observations, which may not be the case in some real-life applications. For example, in the application in \Cref{sec:real-analysis}, typically the tree-based methods are performing better and they are found to select more than 100 variables. Moreover, in such instances, covariates are likely to be correlated and can have grouping effects. Therefore, though the considered penalty-based approaches fail to provide good results in these situations, one may take resort to group LASSO type of estimators which are not evaluated in this work. To that end, it would be a valuable exercise to extend the review and simulations to include methods that incorporate the inherent group structures in the regressor set. 

It is also worth mention that we work with continuous predictors in this paper, and discrete or categorical variables can be added as covariates, although that is not expected to affect the findings significantly. Furthermore, because of computational constraints, we have not incorporated ultra high-dimensional cases in this paper and it would be interesting to see if the performances of the methods alter much in those cases. Naturally, there is a sizeable scope of extending the simulations to more experimental frameworks. 

Regarding the review and comparison of methods, we reiterate that our focus has been restricted on the frequentist approaches in the four typologies. Therefore, Bayesian variable selection procedures such as horseshoe shrinkage priors are not included in this work. Few other recent advances in machine learning (e.g.\ genetic algorithms and its variants) are also not discussed here. A prospective direction to our work would be to cover all such possible techniques and use appropriate measures to compare their virtues across various settings. Last but not the least, we earlier pointed out that there are several methods (CASE, FA-CAR, RSO etc.) the theory of which have been developed for the linear models, but the implementation in generalised linear models are yet to be explored. A future endeavor to improve the existing research on that note would be extremely valuable for both statisticians and practitioners.

\singlespacing
\bibliography{references}

\section{Figures from the simulation study}
\label{sec:figures}

\newpage

\newpage

\newpage

\newpage

\appendix

\section{Additional tables and figures}
\label{sec:appendix}

\setcounter{table}{0}
\renewcommand{\thetable}{A\arabic{table}}
\setcounter{figure}{0}
\renewcommand{\thefigure}{A\arabic{figure}}

\begin{table}[hbt!]
\centering
\caption{performance metrics of the sixteen methods in the low dimensional simulation setting ($n = 100,m = 10,p = 3$).}  
\label{tab:setup11}
\scalebox{0.8}{
\begin{tabular}{lcccccccccc}
 \hline
  & \multicolumn{9}{c}{Independent case} \\
  \hline
Method & Selected & Imp\% & Unimp\% & MSE & MAE & Accuracy & Precision & Recall & Time & Brier \\ 

  \hline
  LASSO & 6.430 & 96.000 & 50.714 & 0.375 & 0.360 & 83.300 & 0.832 & 0.853 & 0.002 & 0.112 \\ 
  ElasticNet & 6.330 & 87.333 & 53.000 & 1.605 & 0.680 & 82.300 & 0.836 & 0.826 & 0.120 & 0.166 \\ 
  ALASSO & 3.910 & 89.667 & 17.429 & 0.210 & 0.265 & 83.900 & 0.839 & 0.858 & 0.003 & 0.110 \\ 
  SparseStep & 2.540 & 77.333 & 3.143 & 0.220 & 0.240 & 84.000 & 0.849 & 0.839 & 0.023 & 0.113 \\ 
  Best Subset & 3.670 & 85.333 & 15.857 & 0.245 & 0.285 & 84.150 & 0.847 & 0.846 & 0.011 & 0.111 \\ 
  \hline
  SCAD & 3.470 & 86.333 & 12.571 & 0.290 & 0.290 & 84.350 & 0.852 & 0.850 & 0.005 & 0.111 \\ 
  MCP & 3.090 & 84.667 & 7.857 & 0.300 & 0.290 & 84.650 & 0.854 & 0.852 & 0.004 & 0.109 \\ 
  \hline
  SIS & 2.710 & 77.667 & 5.429 & 0.385 & 0.335 & 83.600 & 0.852 & 0.827 & 0.001 & 0.113 \\ 
  ISIS & 2.710 & 77.667 & 5.429 & 0.385 & 0.335 & 83.600 & 0.852 & 0.827 & 0.002 & 0.113 \\ 
  SIVS & 3.076 & 90.217 & 9.162 & 0.800 & 0.440 & 84.565 & 0.855 & 0.850 & 0.533 & 0.108 \\ 
  \hline
  varSelRF & 2.670 & 73.000 & 6.857 &  &  & 80.750 & 0.808 & 0.821 & 0.019 & 0.135 \\ 
  PIMP & 2.452 & 75.000 & 2.891 &  &  & 80.893 & 0.807 & 0.813 & 0.057 & 0.134 \\ 
  Boruta & 2.857 & 78.388 & 7.221 &  &  & 81.703 & 0.821 & 0.815 & 0.076 & 0.132 \\ 
  VSURF & 2.765 & 76.871 & 6.560 &  &  & 81.122 & 0.808 & 0.826 & 0.157 & 0.134 \\ 
  RRF & 3.400 & 78.597 & 14.887 &  &  & 80.737 & 0.805 & 0.827 & 0.001 & 0.135 \\ 
  NTA & 3.432 & 79.649 & 14.887 &  &  & 80.368 & 0.809 & 0.801 & 0.006 & 0.133 \\  
   \hline
  & \multicolumn{9}{c}{Correlated case} \\
  \hline
Method & Selected & Imp\% & Unimp\% & MSE & MAE & Accuracy & Precision & Recall & Time & Brier \\ 
  \hline
LASSO & 5.910 & 89.667 & 46.000 & 0.410 & 0.370 & 82.600 & 0.817 & 0.822 & 0.002 & 0.118 \\ 
  ElasticNet & 6.950 & 87.333 & 61.857 & 1.420 & 0.640 & 81.650 & 0.821 & 0.801 & 0.117 & 0.158 \\ 
  ALASSO & 3.750 & 83.333 & 17.857 & 0.300 & 0.310 & 83.750 & 0.829 & 0.836 & 0.003 & 0.115 \\ 
  SparseStep & 2.350 & 69.667 & 3.714 & 0.215 & 0.240 & 83.050 & 0.826 & 0.830 & 0.020 & 0.122 \\ 
  Best Subset & 3.510 & 80.000 & 15.857 & 0.335 & 0.340 & 83.100 & 0.823 & 0.832 & 0.012 & 0.119 \\ 
  \hline
  SCAD & 3.200 & 79.333 & 11.714 & 0.405 & 0.340 & 82.450 & 0.824 & 0.820 & 0.005 & 0.121 \\ 
  MCP & 2.780 & 77.000 & 6.714 & 0.350 & 0.310 & 82.800 & 0.826 & 0.826 & 0.005 & 0.118 \\ 
  \hline
  SIS & 2.450 & 70.333 & 4.857 & 0.525 & 0.380 & 82.050 & 0.828 & 0.813 & 0.001 & 0.130 \\ 
  ISIS & 2.450 & 70.333 & 4.857 & 0.525 & 0.380 & 82.050 & 0.828 & 0.813 & 0.002 & 0.130 \\ 
  SIVS & 2.955 & 86.743 & 11.364 & 0.865 & 0.480 & 83.920 & 0.833 & 0.839 & 0.447 & 0.113 \\ 
  \hline
  varSelRF & 2.640 & 66.000 & 9.429 &  &  & 77.650 & 0.780 & 0.758 & 0.017 & 0.153 \\ 
  PIMP & 2.852 & 71.970 & 9.903 &  &  & 77.386 & 0.773 & 0.739 & 0.058 & 0.153 \\ 
  Boruta & 3.905 & 76.842 & 22.857 &  &  & 79.105 & 0.793 & 0.775 & 0.095 & 0.148 \\ 
  VSURF & 2.753 & 68.313 & 10.053 &  &  & 78.148 & 0.790 & 0.761 & 0.156 & 0.151 \\ 
  RRF & 3.742 & 74.571 & 21.502 &  &  & 78.763 & 0.788 & 0.773 & 0.001 & 0.150 \\ 
  NTA & 4.789 & 79.649 & 34.286 &  &  & 78.579 & 0.791 & 0.773 & 0.006 & 0.147 \\
   \hline
\end{tabular}}
\end{table}

\begin{table}[hbt!]
\centering
\caption{performance metrics of the sixteen methods in the low dimensional simulation setting ($n = 200,m = 10,p = 3$).}  
\label{tab:setup14}
\scalebox{0.8}{
\begin{tabular}{lcccccccccc}
 \hline
  & \multicolumn{9}{c}{Independent case} \\
  \hline
Method & Selected & Imp\% & Unimp\% & MSE & MAE & Accuracy & Precision & Recall & Time & Brier \\ 

  \hline
  LASSO & 6.550 & 98.333 & 51.429 & 0.205 & 0.280 & 85.275 & 0.850 & 0.848 & 0.002 & 0.104 \\ 
  ElasticNet & 6.760 & 94.333 & 56.143 & 0.785 & 0.475 & 84.525 & 0.842 & 0.842 & 0.128 & 0.144 \\ 
  ALASSO & 3.840 & 95.000 & 14.143 & 0.140 & 0.205 & 85.225 & 0.848 & 0.846 & 0.003 & 0.102 \\ 
  SparseStep & 3.140 & 88.333 & 7.000 & 0.100 & 0.165 & 85.500 & 0.847 & 0.855 & 0.006 & 0.102 \\ 
  Best Subset & 3.830 & 93.000 & 14.857 & 0.145 & 0.215 & 85.050 & 0.843 & 0.851 & 0.014 & 0.102 \\ 
  \hline
  SCAD & 3.720 & 93.667 & 13.000 & 0.115 & 0.185 & 85.150 & 0.844 & 0.853 & 0.006 & 0.104 \\ 
  MCP & 3.370 & 91.667 & 8.857 & 0.110 & 0.180 & 85.400 & 0.846 & 0.856 & 0.004 & 0.102 \\ 
  \hline
  SIS & 2.780 & 85.667 & 3.000 & 0.150 & 0.200 & 85.425 & 0.847 & 0.855 & 0.001 & 0.104 \\ 
  ISIS & 2.780 & 85.667 & 3.000 & 0.150 & 0.200 & 85.425 & 0.847 & 0.855 & 0.002 & 0.104 \\ 
  SIVS & 3.061 & 95.238 & 6.851 & 0.620 & 0.380 & 85.561 & 0.846 & 0.859 & 0.531 & 0.101 \\ 
  \hline
  varSelRF & 2.900 & 80.334 & 7.000 &  &  & 82.525 & 0.822 & 0.821 & 0.028 & 0.126 \\ 
  PIMP & 2.556 & 80.741 & 1.905 &  &  & 82.667 & 0.824 & 0.819 & 0.136 & 0.125 \\ 
  Boruta & 3.040 & 84.512 & 7.215 &  &  & 82.475 & 0.821 & 0.819 & 0.140 & 0.126 \\ 
  VSURF & 3.051 & 82.492 & 8.225 &  &  & 82.551 & 0.821 & 0.820 & 0.247 & 0.127 \\ 
  RRF & 3.347 & 79.252 & 13.849 &  &  & 82.270 & 0.820 & 0.818 & 0.003 & 0.128 \\ 
  NTA & 3.586 & 85.522 & 14.574 &  &  & 82.828 & 0.822 & 0.825 & 0.012 & 0.125 \\ 
   \hline
  & \multicolumn{9}{c}{Correlated case} \\
  \hline
Method & Selected & Imp\% & Unimp\% & MSE & MAE & Accuracy & Precision & Recall & Time & Brier \\ 
  \hline
LASSO & 6.510 & 98.667 & 50.714 & 0.205 & 0.275 & 84.775 & 0.842 & 0.843 & 0.002 & 0.109 \\ 
  ElasticNet & 7.740 & 93.000 & 70.714 & 0.375 & 0.430 & 83.625 & 0.829 & 0.837 & 0.139 & 0.136 \\ 
  ALASSO & 3.930 & 93.667 & 16.000 & 0.130 & 0.200 & 84.625 & 0.839 & 0.843 & 0.004 & 0.108 \\ 
  SparseStep & 2.900 & 85.333 & 4.857 & 0.090 & 0.170 & 84.100 & 0.837 & 0.838 & 0.008 & 0.111 \\ 
  Best Subset & 3.570 & 89.667 & 12.571 & 0.110 & 0.190 & 84.100 & 0.835 & 0.839 & 0.016 & 0.112 \\ 
  \hline
  SCAD & 3.380 & 88.667 & 10.286 & 0.145 & 0.210 & 84.225 & 0.833 & 0.846 & 0.009 & 0.112 \\ 
  MCP & 3.140 & 87.667 & 7.286 & 0.120 & 0.180 & 84.500 & 0.840 & 0.841 & 0.007 & 0.110 \\ 
  \hline
  SIS & 2.680 & 83.000 & 2.714 & 0.150 & 0.200 & 84.050 & 0.834 & 0.839 & 0.001 & 0.112 \\ 
  ISIS & 2.680 & 83.000 & 2.714 & 0.150 & 0.200 & 84.050 & 0.834 & 0.839 & 0.003 & 0.112 \\ 
  SIVS & 2.918 & 95.238 & 6.123 & 0.610 & 0.375 & 84.617 & 0.842 & 0.841 & 0.465 & 0.108 \\
  \hline
  varSelRF & 2.740 & 74.334 & 7.286 &  &  & 81.650 & 0.801 & 0.829 & 0.028 & 0.134 \\ 
  PIMP & 3.327 & 78.912 & 13.703 &  &  & 81.276 & 0.800 & 0.825 & 0.152 & 0.134 \\ 
  Boruta & 4.636 & 85.859 & 29.437 &  &  & 81.313 & 0.799 & 0.822 & 0.196 & 0.133 \\ 
  VSURF & 2.948 & 75.695 & 9.673 &  &  & 81.615 & 0.802 & 0.828 & 0.277 & 0.132 \\ 
  RRF & 3.778 & 81.482 & 19.048 &  &  & 81.263 & 0.799 & 0.818 & 0.003 & 0.132 \\ 
  NTA & 5.670 & 89.773 & 42.532 &  &  & 80.994 & 0.797 & 0.817 & 0.013 & 0.136 \\ 
   \hline
\end{tabular}}
\end{table}

\begin{table}[hbt!]
\centering
\caption{performance metrics of the sixteen methods in the low dimensional simulation setting ($n = 500,m = 10,p = 3$).}  
\label{tab:setup17}
\scalebox{0.8}{
\begin{tabular}{lcccccccccc}
 \hline
  & \multicolumn{9}{c}{Independent case} \\
  \hline
Method & Selected & Imp\% & Unimp\% & MSE & MAE & Accuracy & Precision & Recall & Time & Brier \\ 

  \hline
  LASSO & 7.350 & 100.000 & 62.143 & 0.120 & 0.210 & 84.320 & 0.841 & 0.852 & 0.002 & 0.107 \\ 
  ElasticNet & 6.780 & 97.667 & 55.000 & 0.300 & 0.295 & 83.910 & 0.837 & 0.852 & 0.161 & 0.139 \\ 
  ALASSO & 3.970 & 99.000 & 14.286 & 0.065 & 0.130 & 84.590 & 0.844 & 0.853 & 0.005 & 0.106 \\ 
  SparseStep & 3.540 & 97.333 & 8.857 & 0.040 & 0.100 & 84.320 & 0.840 & 0.855 & 0.015 & 0.107 \\ 
  Best Subset & 3.610 & 97.667 & 9.714 & 0.060 & 0.130 & 84.280 & 0.840 & 0.855 & 0.087 & 0.107 \\ 
  \hline
  SCAD & 3.590 & 92.333 & 11.714 & 0.060 & 0.130 & 83.980 & 0.838 & 0.849 & 0.017 & 0.111 \\ 
  MCP & 3.330 & 94.333 & 7.143 & 0.050 & 0.120 & 84.320 & 0.840 & 0.854 & 0.009 & 0.108 \\ 
  \hline
  SIS & 2.770 & 90.333 & 0.857 & 0.060 & 0.120 & 84.230 & 0.840 & 0.852 & 0.002 & 0.110 \\ 
  ISIS & 2.770 & 90.333 & 0.857 & 0.060 & 0.120 & 84.230 & 0.840 & 0.852 & 0.005 & 0.110 \\ 
  SIVS & 3.050 & 99.667 & 2.429 & 0.365 & 0.310 & 84.540 & 0.843 & 0.856 & 0.697 & 0.106 \\ 
  \hline
  varSelRF & 3.090 & 90.000 & 5.571 &  &  & 81.560 & 0.819 & 0.818 & 0.058 & 0.129 \\ 
  PIMP & 2.796 & 88.095 & 2.187 &  &  & 81.429 & 0.816 & 0.821 & 0.423 & 0.130 \\ 
  Boruta & 3.354 & 95.623 & 6.926 &  &  & 81.838 & 0.822 & 0.822 & 0.313 & 0.127 \\ 
  VSURF & 3.520 & 92.000 & 10.857 &  &  & 81.730 & 0.819 & 0.822 & 0.644 & 0.127 \\ 
  RRF & 3.560 & 83.667 & 15.000 &  &  & 81.550 & 0.817 & 0.821 & 0.008 & 0.129 \\ 
  NTA & 4.082 & 95.918 & 17.201 &  &  & 82.051 & 0.824 & 0.825 & 0.030 & 0.127 \\ 
   \hline
  & \multicolumn{9}{c}{Correlated case} \\
  \hline
Method & Selected & Imp\% & Unimp\% & MSE & MAE & Accuracy & Precision & Recall & Time & Brier \\ 
  \hline
  LASSO & 7.110 & 99.333 & 59.000 & 0.130 & 0.210 & 84.850 & 0.850 & 0.849 & 0.002 & 0.105 \\ 
  ElasticNet & 9.000 & 97.667 & 86.714 & 0.080 & 0.230 & 84.390 & 0.843 & 0.849 & 0.164 & 0.119 \\ 
  ALASSO & 4.240 & 98.000 & 18.571 & 0.075 & 0.160 & 84.900 & 0.849 & 0.850 & 0.005 & 0.105 \\ 
  SparseStep & 3.450 & 95.667 & 8.286 & 0.050 & 0.115 & 84.870 & 0.849 & 0.851 & 0.017 & 0.105 \\ 
  Best Subset & 3.530 & 96.333 & 9.143 & 0.070 & 0.140 & 84.980 & 0.850 & 0.851 & 0.093 & 0.105 \\ 
  \hline
  SCAD & 3.600 & 92.000 & 12.000 & 0.075 & 0.150 & 84.350 & 0.844 & 0.845 & 0.019 & 0.109 \\ 
  MCP & 3.400 & 93.000 & 8.714 & 0.050 & 0.120 & 84.810 & 0.848 & 0.851 & 0.012 & 0.106 \\ 
  \hline
  SIS & 2.830 & 90.000 & 1.857 & 0.070 & 0.140 & 84.620 & 0.847 & 0.847 & 0.002 & 0.108 \\ 
  ISIS & 2.830 & 90.000 & 1.857 & 0.070 & 0.140 & 84.620 & 0.847 & 0.847 & 0.006 & 0.108 \\ 
  SIVS & 3.020 & 98.317 & 3.175 & 0.420 & 0.320 & 84.758 & 0.847 & 0.850 & 0.569 & 0.105 \\ 
  \hline
  varSelRF & 3.030 & 79.334 & 9.286 &  &  & 82.270 & 0.821 & 0.827 & 0.056 & 0.126 \\ 
  PIMP & 4.460 & 85.667 & 27.000 &  &  & 82.470 & 0.826 & 0.826 & 0.426 & 0.124 \\ 
  Boruta & 6.121 & 93.266 & 47.475 &  &  & 82.677 & 0.825 & 0.832 & 0.500 & 0.123 \\ 
  VSURF & 3.381 & 86.254 & 11.340 &  &  & 82.938 & 0.827 & 0.833 & 0.596 & 0.121 \\ 
  RRF & 3.899 & 79.798 & 21.501 &  &  & 82.061 & 0.819 & 0.826 & 0.008 & 0.126 \\ 
  NTA & 7.274 & 95.161 & 63.134 &  &  & 81.823 & 0.813 & 0.826 & 0.030 & 0.129 \\ 
   \hline
\end{tabular}}
\end{table}

\begin{table}[!hbt]
\centering
\caption{performance metrics of the sixteen methods in the moderate dimensional simulation setting ($n = 100,m = 100,p = 5$).}  
\label{tab:setup12}
\scalebox{0.8}{
\begin{tabular}{lcccccccccc}
  \hline
  & \multicolumn{9}{c}{Independent case} \\
  \hline
Method & Selected & Imp\% & Unimp\% & MSE & MAE & Accuracy & Precision & Recall & Time & Brier \\ 
  \hline
LASSO & 18.430 & 81.400 & 15.116 & 0.190 & 0.110 & 83.450 & 0.819 & 0.834 & 0.002 & 0.123 \\ 
  ElasticNet & 17.630 & 70.600 & 14.842 & 0.370 & 0.140 & 79.650 & 0.786 & 0.816 & 0.138 & 0.189 \\ 
  ALASSO & 13.140 & 75.200 & 9.874 & 0.110 & 0.090 & 83.250 & 0.822 & 0.825 & 0.004 & 0.120 \\ 
  SparseStep & 2.610 & 50.400 & 0.095 & 0.095 & 0.060 & 82.350 & 0.808 & 0.827 & 0.067 & 0.126 \\ 
  Best Subset & 3.850 & 60.800 & 0.853 & 0.110 & 0.070 & 82.850 & 0.819 & 0.827 & 0.048 & 0.119 \\ 
  \hline
  SCAD & 9.950 & 77.400 & 6.400 & 0.130 & 0.080 & 84.200 & 0.833 & 0.839 & 0.012 & 0.111 \\ 
  MCP & 5.780 & 69.400 & 2.432 & 0.140 & 0.080 & 84.000 & 0.828 & 0.836 & 0.010 & 0.112 \\ 
  \hline
  SIS & 3.350 & 51.200 & 0.832 & 0.125 & 0.070 & 82.350 & 0.816 & 0.822 & 0.001 & 0.127 \\ 
  ISIS & 3.850 & 62.800 & 0.747 & 0.070 & 0.060 & 82.600 & 0.814 & 0.830 & 0.003 & 0.126 \\ 
  SIVS & 4.060 & 60.800 & 1.116 & 0.195 & 0.090 & 82.550 & 0.809 & 0.829 & 1.429 & 0.124 \\ 
  \hline
  varSelRF & 6.450 & 54.000 & 3.947 &  &  & 76.500 & 0.745 & 0.783 & 0.049 & 0.164 \\ 
  PIMP & 7.890 & 54.800 & 5.421 &  &  & 75.750 & 0.753 & 0.762 & 0.186 & 0.165 \\ 
  Boruta & 5.677 & 53.737 & 3.147 &  &  & 75.505 & 0.738 & 0.760 & 0.143 & 0.164 \\ 
  VSURF & 4.374 & 49.697 & 1.988 &  &  & 76.768 & 0.756 & 0.767 & 0.286 & 0.162 \\ 
  RRF & 3.670 & 47.000 & 1.390 &  &  & 76.650 & 0.750 & 0.771 & 0.004 & 0.164 \\ 
  NTA & 12.360 & 62.600 & 9.716 &  &  & 74.600 & 0.729 & 0.752 & 0.032 & 0.176 \\ 
   \hline
  & \multicolumn{9}{c}{Correlated case} \\
  \hline
Method & Selected & Imp\% & Unimp\% & MSE & MAE & Accuracy & Precision & Recall & Time & Brier \\ 
  \hline
LASSO & 19.200 & 77.400 & 16.137 & 0.190 & 0.110 & 81.000 & 0.802 & 0.816 & 0.002 & 0.132 \\ 
  ElasticNet & 21.890 & 67.600 & 19.484 & 0.365 & 0.140 & 78.050 & 0.787 & 0.785 & 0.133 & 0.187 \\ 
  ALASSO & 12.450 & 70.400 & 9.400 & 0.120 & 0.090 & 81.600 & 0.809 & 0.822 & 0.004 & 0.130 \\ 
  SparseStep & 2.600 & 48.600 & 0.179 & 0.110 & 0.070 & 80.900 & 0.811 & 0.809 & 0.064 & 0.135 \\ 
  Best Subset & 5.400 & 60.600 & 2.495 & 0.130 & 0.085 & 81.500 & 0.809 & 0.823 & 0.050 & 0.134 \\ 
  \hline
  SCAD & 9.900 & 73.800 & 6.537 & 0.130 & 0.080 & 82.800 & 0.838 & 0.823 & 0.012 & 0.122 \\ 
  MCP & 5.890 & 67.000 & 2.674 & 0.125 & 0.075 & 82.650 & 0.829 & 0.830 & 0.010 & 0.119 \\ 
  \hline
  SIS & 2.610 & 40.400 & 0.621 & 0.170 & 0.090 & 77.300 & 0.776 & 0.770 & 0.001 & 0.152 \\ 
  ISIS & 3.810 & 56.000 & 1.063 & 0.135 & 0.080 & 81.400 & 0.818 & 0.808 & 0.003 & 0.143 \\ 
  SIVS & 3.930 & 60.800 & 0.937 & 0.200 & 0.090 & 82.150 & 0.822 & 0.827 & 1.272 & 0.124 \\ 
  \hline
  varSelRF & 5.440 & 46.600 & 3.274 &  &  & 74.800 & 0.752 & 0.749 & 0.045 & 0.172 \\ 
  PIMP & 8.890 & 50.200 & 6.716 &  &  & 73.700 & 0.737 & 0.738 & 0.187 & 0.177 \\ 
  Boruta & 7.040 & 50.800 & 4.737 &  &  & 73.850 & 0.737 & 0.741 & 0.200 & 0.178 \\ 
  VSURF & 4.102 & 44.490 & 1.976 &  &  & 74.847 & 0.749 & 0.757 & 0.283 & 0.176 \\ 
  RRF & 3.960 & 42.800 & 1.916 &  &  & 74.350 & 0.737 & 0.755 & 0.004 & 0.177 \\ 
  NTA & 14.510 & 59.000 & 12.168 &  &  & 74.200 & 0.738 & 0.741 & 0.032 & 0.180 \\ 
   \hline
\end{tabular}}
\end{table}

\begin{table}[!hbt]
\centering
\caption{performance metrics of the sixteen methods in the moderate dimensional simulation setting ($n = 200,m = 100,p = 5$).}  
\label{tab:setup15}
\scalebox{0.8}{
\begin{tabular}{lcccccccccc}
  \hline
  & \multicolumn{9}{c}{Independent case} \\
  \hline
Method & Selected & Imp\% & Unimp\% & MSE & MAE & Accuracy & Precision & Recall & Time & Brier \\ 
  \hline
LASSO & 24.610 & 91.600 & 21.084 & 0.100 & 0.080 & 85.600 & 0.854 & 0.854 & 0.005 & 0.107 \\ 
  ElasticNet & 15.900 & 75.200 & 12.779 & 0.300 & 0.100 & 82.800 & 0.836 & 0.832 & 0.249 & 0.172 \\ 
  ALASSO & 16.260 & 88.000 & 12.484 & 0.045 & 0.070 & 85.175 & 0.851 & 0.846 & 0.009 & 0.103 \\ 
  SparseStep & 3.450 & 66.800 & 0.116 & 0.030 & 0.035 & 85.925 & 0.861 & 0.857 & 0.082 & 0.097 \\ 
  Best Subset & 4.190 & 70.200 & 0.716 & 0.050 & 0.040 & 85.525 & 0.860 & 0.847 & 0.060 & 0.100 \\ 
  \hline
  SCAD & 9.580 & 83.200 & 5.705 & 0.050 & 0.050 & 86.100 & 0.865 & 0.856 & 0.026 & 0.097 \\ 
  MCP & 6.190 & 80.200 & 2.295 & 0.040 & 0.050 & 86.475 & 0.867 & 0.861 & 0.022 & 0.094 \\ 
  \hline
  SIS & 7.360 & 75.000 & 3.800 & 0.230 & 0.090 & 84.175 & 0.843 & 0.839 & 0.004 & 0.131 \\ 
  ISIS & 7.360 & 75.000 & 3.800 & 0.230 & 0.090 & 84.175 & 0.843 & 0.839 & 0.015 & 0.131 \\ 
  SIVS & 4.610 & 72.600 & 1.032 & 0.120 & 0.070 & 85.550 & 0.857 & 0.852 & 3.582 & 0.101 \\ 
  \hline
  varSelRF & 8.000 & 65.000 & 5.000 &  &  & 80.800 & 0.809 & 0.811 & 0.083 & 0.141 \\ 
  PIMP & 8.300 & 65.000 & 5.316 &  &  & 80.600 & 0.809 & 0.806 & 0.464 & 0.139 \\ 
  Boruta & 6.140 & 64.200 & 3.084 &  &  & 81.825 & 0.819 & 0.820 & 0.307 & 0.137 \\ 
  VSURF & 4.890 & 59.800 & 2.000 &  &  & 81.200 & 0.812 & 0.814 & 0.557 & 0.137 \\ 
  RRF & 3.302 & 50.833 & 0.801 &  &  & 80.182 & 0.801 & 0.807 & 0.009 & 0.141 \\ 
  NTA & 12.690 & 71.600 & 9.589 &  &  & 79.750 & 0.801 & 0.796 & 0.061 & 0.149 \\ 
   \hline
  & \multicolumn{9}{c}{Correlated case} \\
  \hline
Method & Selected & Imp\% & Unimp\% & MSE & MAE & Accuracy & Precision & Recall & Time & Brier \\ 
  \hline
LASSO & 24.310 & 91.400 & 20.779 & 0.110 & 0.085 & 85.625 & 0.862 & 0.857 & 0.006 & 0.100 \\ 
  ElasticNet & 19.340 & 77.000 & 16.305 & 0.310 & 0.110 & 82.100 & 0.827 & 0.839 & 0.266 & 0.164 \\ 
  ALASSO & 16.140 & 86.600 & 12.432 & 0.060 & 0.070 & 85.850 & 0.867 & 0.855 & 0.011 & 0.099 \\ 
  SparseStep & 3.460 & 64.800 & 0.232 & 0.030 & 0.040 & 86.775 & 0.871 & 0.872 & 0.082 & 0.094 \\ 
  Best Subset & 4.260 & 70.200 & 0.790 & 0.040 & 0.050 & 86.500 & 0.868 & 0.866 & 0.062 & 0.094 \\ 
  \hline
  SCAD & 9.400 & 84.000 & 5.474 & 0.050 & 0.050 & 87.125 & 0.874 & 0.874 & 0.027 & 0.091 \\ 
  MCP & 6.240 & 79.000 & 2.411 & 0.040 & 0.050 & 86.975 & 0.872 & 0.875 & 0.023 & 0.092 \\ 
  \hline
  SIS & 6.940 & 71.600 & 3.537 & 0.185 & 0.090 & 85.600 & 0.856 & 0.863 & 0.004 & 0.121 \\ 
  ISIS & 6.940 & 71.600 & 3.537 & 0.185 & 0.090 & 85.600 & 0.856 & 0.863 & 0.016 & 0.121 \\ 
  SIVS & 4.380 & 71.000 & 0.874 & 0.125 & 0.070 & 87.100 & 0.876 & 0.871 & 3.556 & 0.093 \\ 
  \hline
  varSelRF & 6.820 & 59.400 & 4.053 &  &  & 81.825 & 0.820 & 0.834 & 0.088 & 0.135 \\ 
  PIMP & 10.400 & 62.800 & 7.642 &  &  & 81.000 & 0.815 & 0.820 & 0.479 & 0.142 \\ 
  Boruta & 8.420 & 64.600 & 5.463 &  &  & 81.450 & 0.817 & 0.829 & 0.403 & 0.136 \\ 
  VSURF & 4.838 & 58.586 & 2.010 &  &  & 81.338 & 0.817 & 0.826 & 0.635 & 0.134 \\ 
  RRF & 3.570 & 48.200 & 1.221 &  &  & 80.900 & 0.812 & 0.822 & 0.009 & 0.139 \\ 
  NTA & 16.950 & 72.200 & 14.042 &  &  & 79.950 & 0.805 & 0.811 & 0.063 & 0.149 \\ 
   \hline
\end{tabular}}
\end{table}

\begin{table}[!hbt]
\centering
\caption{performance metrics of the sixteen methods in the moderate dimensional simulation setting ($n = 500,m = 100,p = 5$).}  
\label{tab:setup18}
\scalebox{0.8}{
\begin{tabular}{lcccccccccc}
  \hline
  & \multicolumn{9}{c}{Independent case} \\
  \hline
Method & Selected & Imp\% & Unimp\% & MSE & MAE & Accuracy & Precision & Recall & Time & Brier \\ 
  \hline
LASSO & 32.400 & 98.800 & 28.905 & 0.075 & 0.080 & 87.230 & 0.873 & 0.870 & 0.049 & 0.090 \\ 
  ElasticNet & 14.900 & 88.800 & 11.011 & 0.325 & 0.110 & 86.110 & 0.867 & 0.857 & 1.208 & 0.161 \\ 
  ALASSO & 18.550 & 97.200 & 14.411 & 0.020 & 0.050 & 87.230 & 0.873 & 0.870 & 0.032 & 0.090 \\ 
  SparseStep & 4.670 & 88.000 & 0.284 & 0.010 & 0.020 & 87.980 & 0.884 & 0.875 & 0.135 & 0.085 \\ 
  Best Subset & 4.870 & 88.200 & 0.484 & 0.020 & 0.030 & 87.900 & 0.882 & 0.874 & 0.233 & 0.085 \\ 
  \hline
  SCAD & 10.110 & 93.000 & 5.747 & 0.010 & 0.030 & 88.010 & 0.881 & 0.878 & 0.111 & 0.087 \\ 
  MCP & 7.010 & 92.000 & 2.537 & 0.010 & 0.030 & 88.190 & 0.884 & 0.878 & 0.101 & 0.085 \\ 
  \hline
  SIS & 4.770 & 87.400 & 0.421 & 0.030 & 0.040 & 87.340 & 0.876 & 0.869 & 0.013 & 0.089 \\ 
  ISIS & 4.770 & 87.400 & 0.421 & 0.030 & 0.040 & 87.340 & 0.876 & 0.869 & 0.025 & 0.089 \\ 
  SIVS & 4.152 & 74.949 & 0.425 & 0.130 & 0.070 & 86.394 & 0.866 & 0.859 & 12.917 & 0.095 \\ 
  \hline
  varSelRF & 7.300 & 77.800 & 3.589 &  &  & 84.070 & 0.844 & 0.836 & 0.245 & 0.119 \\ 
  PIMP & 8.830 & 78.000 & 5.189 &  &  & 84.300 & 0.846 & 0.839 & 1.825 & 0.120 \\ 
  Boruta & 5.920 & 79.400 & 2.053 &  &  & 83.990 & 0.843 & 0.835 & 1.030 & 0.116 \\ 
  VSURF & 5.330 & 75.000 & 1.663 &  &  & 84.200 & 0.844 & 0.840 & 1.656 & 0.115 \\ 
  RRF & 3.469 & 56.939 & 0.655 &  &  & 82.122 & 0.824 & 0.817 & 0.031 & 0.125 \\ 
  NTA & 12.990 & 83.800 & 9.263 &  &  & 83.950 & 0.842 & 0.836 & 0.169 & 0.127 \\ 
   \hline
  & \multicolumn{9}{c}{Correlated case} \\
  \hline
Method & Selected & Imp\% & Unimp\% & MSE & MAE & Accuracy & Precision & Recall & Time & Brier \\ 
  \hline
LASSO & 29.450 & 98.600 & 25.810 & 0.070 & 0.080 & 87.970 & 0.875 & 0.885 & 0.050 & 0.087 \\ 
  ElasticNet & 17.630 & 83.800 & 14.147 & 0.330 & 0.110 & 85.120 & 0.844 & 0.864 & 1.134 & 0.156 \\ 
  ALASSO & 17.840 & 96.400 & 13.705 & 0.020 & 0.050 & 87.990 & 0.875 & 0.884 & 0.026 & 0.085 \\ 
  SparseStep & 4.610 & 86.800 & 0.284 & 0.010 & 0.020 & 87.990 & 0.876 & 0.884 & 0.004 & 0.084 \\ 
  Best Subset & 4.900 & 88.200 & 0.516 & 0.020 & 0.030 & 87.990 & 0.877 & 0.884 & 0.009 & 0.085 \\ 
  \hline
  SCAD & 9.460 & 93.800 & 5.021 & 0.020 & 0.030 & 87.900 & 0.874 & 0.884 & 0.074 & 0.085 \\ 
  MCP & 6.440 & 92.400 & 1.916 & 0.010 & 0.030 & 87.920 & 0.874 & 0.885 & 0.068 & 0.083 \\ 
  \hline
  SIS & 4.800 & 87.800 & 0.432 & 0.040 & 0.040 & 87.650 & 0.871 & 0.884 & 0.009 & 0.087 \\ 
  ISIS & 4.800 & 87.800 & 0.432 & 0.040 & 0.040 & 87.650 & 0.871 & 0.884 & 0.017 & 0.087 \\ 
  SIVS & 4.200 & 74.400 & 0.505 & 0.120 & 0.070 & 86.640 & 0.863 & 0.872 & 10.134 & 0.093 \\ 
  \hline
  varSelRF & 7.300 & 71.800 & 3.905 &  &  & 83.800 & 0.835 & 0.841 & 0.252 & 0.119 \\ 
  PIMP & 12.920 & 77.400 & 9.526 &  &  & 83.650 & 0.832 & 0.842 & 2.059 & 0.126 \\ 
  Boruta & 10.130 & 78.600 & 6.526 &  &  & 84.150 & 0.837 & 0.848 & 1.020 & 0.121 \\ 
  VSURF & 4.880 & 72.800 & 1.305 &  &  & 84.190 & 0.836 & 0.849 & 1.828 & 0.113 \\ 
  RRF & 3.680 & 55.400 & 0.958 &  &  & 81.830 & 0.812 & 0.827 & 0.032 & 0.127 \\ 
  NTA & 20.110 & 84.400 & 16.726 &  &  & 82.830 & 0.822 & 0.837 & 0.179 & 0.134 \\ 
   \hline
\end{tabular}}
\end{table}

\begin{table}[hbt!]
\centering
\caption{performance metrics of the sixteen methods in the high dimensional simulation setting ($n = 100,m = 1000,p = 10$).}  
\label{tab:setup13}
\scalebox{0.8}{
\begin{tabular}{lcccccccccc}
   \hline
  & \multicolumn{9}{c}{Independent case} \\
  \hline
Method & Selected & Imp\% & Unimp\% & MSE & MAE & Accuracy & Precision & Recall & Time & Brier \\ 
  \hline
LASSO & 24.310 & 40.700 & 2.044 & 0.060 & 0.020 & 74.150 & 0.740 & 0.763 & 0.005 & 0.179 \\ 
  ElasticNet & 53.540 & 38.600 & 5.018 & 0.070 & 0.030 & 69.800 & 0.722 & 0.663 & 0.277 & 0.219 \\ 
  ALASSO & 30.540 & 43.800 & 2.642 & 0.050 & 0.030 & 72.750 & 0.729 & 0.740 & 0.016 & 0.194 \\ 
  SparseStep & 1.800 & 15.900 & 0.021 & 0.050 & 0.020 & 69.650 & 0.708 & 0.700 & 0.075 & 0.196 \\ 
  Best Subset & 4.930 & 24.700 & 0.248 & 0.060 & 0.020 & 71.700 & 0.736 & 0.700 & 0.066 & 0.182 \\ 
  \hline
  SCAD & 19.350 & 40.200 & 1.548 & 0.060 & 0.020 & 74.300 & 0.764 & 0.722 & 0.019 & 0.174 \\ 
  MCP & 8.010 & 32.300 & 0.483 & 0.060 & 0.020 & 74.950 & 0.769 & 0.746 & 0.014 & 0.169 \\ 
  \hline
  SIS & 3.810 & 21.300 & 0.170 & 0.050 & 0.020 & 70.700 & 0.714 & 0.697 & 0.001 & 0.208 \\ 
  ISIS & 3.910 & 24.300 & 0.149 & 0.050 & 0.020 & 71.900 & 0.728 & 0.716 & 0.011 & 0.215 \\ 
  SIVS & 4.190 & 26.400 & 0.157 & 0.060 & 0.020 & 72.100 & 0.726 & 0.721 & 2.115 & 0.192 \\ 
  \hline
  varSelRF & 42.420 & 34.400 & 3.937 &  &  & 63.850 & 0.654 & 0.601 & 0.214 & 0.224 \\ 
  PIMP & 79.520 & 33.200 & 7.697 &  &  & 61.950 & 0.642 & 0.548 & 1.325 & 0.232 \\ 
  Boruta & 8.460 & 21.700 & 0.635 &  &  & 64.150 & 0.650 & 0.618 & 0.312 & 0.221 \\ 
  VSURF & 8.550 & 20.800 & 0.653 &  &  & 63.550 & 0.638 & 0.627 & 1.302 & 0.227 \\ 
  RRF & 5.220 & 19.400 & 0.331 &  &  & 67.250 & 0.682 & 0.653 & 0.029 & 0.215 \\ 
  NTA & 77.850 & 37.700 & 7.483 &  &  & 62.550 & 0.654 & 0.584 & 0.265 & 0.230 \\  
 \hline
  & \multicolumn{9}{c}{Correlated case} \\
  \hline
Method & Selected & Imp\% & Unimp\% & MSE & MAE & Accuracy & Precision & Recall & Time & Brier \\ 
  \hline
LASSO & 26.220 & 40.300 & 2.241 & 0.065 & 0.020 & 76.350 & 0.767 & 0.773 & 0.004 & 0.167 \\ 
  ElasticNet & 63.300 & 39.300 & 5.997 & 0.070 & 0.030 & 71.100 & 0.739 & 0.693 & 0.289 & 0.221 \\ 
  ALASSO & 29.700 & 42.000 & 2.575 & 0.050 & 0.030 & 73.800 & 0.743 & 0.734 & 0.017 & 0.181 \\ 
  SparseStep & 1.590 & 14.200 & 0.017 & 0.050 & 0.020 & 71.500 & 0.736 & 0.697 & 0.078 & 0.187 \\ 
  Best Subset & 5.240 & 24.800 & 0.279 & 0.060 & 0.020 & 73.100 & 0.751 & 0.716 & 0.069 & 0.175 \\ 
  \hline
  SCAD & 20.260 & 39.400 & 1.648 & 0.060 & 0.020 & 75.550 & 0.781 & 0.738 & 0.020 & 0.162 \\ 
  MCP & 8.060 & 31.400 & 0.497 & 0.060 & 0.020 & 75.850 & 0.788 & 0.737 & 0.015 & 0.160 \\ 
  \hline
  SIS & 3.550 & 20.300 & 0.154 & 0.050 & 0.020 & 73.100 & 0.738 & 0.724 & 0.001 & 0.188 \\ 
  ISIS & 3.970 & 24.400 & 0.155 & 0.040 & 0.020 & 73.650 & 0.742 & 0.728 & 0.011 & 0.191 \\ 
  SIVS & 4.111 & 26.061 & 0.152 & 0.060 & 0.020 & 75.909 & 0.761 & 0.759 & 1.954 & 0.168 \\ 
  \hline
  varSelRF & 35.160 & 32.500 & 3.223 &  &  & 67.950 & 0.703 & 0.674 & 0.222 & 0.215 \\ 
  PIMP & 80.730 & 33.200 & 7.819 &  &  & 63.550 & 0.686 & 0.600 & 1.376 & 0.226 \\ 
  Boruta & 9.920 & 22.900 & 0.771 &  &  & 68.200 & 0.699 & 0.666 & 0.339 & 0.208 \\ 
  VSURF & 7.940 & 21.100 & 0.589 &  &  & 66.950 & 0.691 & 0.642 & 1.310 & 0.216 \\ 
  RRF & 5.200 & 18.300 & 0.340 &  &  & 67.250 & 0.677 & 0.670 & 0.030 & 0.213 \\ 
  NTA & 82.590 & 39.800 & 7.941 &  &  & 63.750 & 0.688 & 0.611 & 0.277 & 0.226 \\ 
   \hline
\end{tabular}}
\end{table}

\begin{table}[hbt!]
\centering
\caption{performance metrics of the sixteen methods in the high dimensional simulation setting ($n = 200,m = 1000,p = 10$).}  
\label{tab:setup16}
\scalebox{0.8}{
\begin{tabular}{lcccccccccc}
   \hline
  & \multicolumn{9}{c}{Independent case} \\
  \hline
Method & Selected & Imp\% & Unimp\% & MSE & MAE & Accuracy & Precision & Recall & Time & Brier \\ 
  \hline
LASSO & 53.700 & 68.600 & 4.731 & 0.060 & 0.020 & 82.700 & 0.827 & 0.832 & 0.009 & 0.121 \\ 
  ElasticNet & 47.250 & 57.600 & 4.191 & 0.050 & 0.030 & 80.275 & 0.802 & 0.817 & 0.500 & 0.177 \\ 
  ALASSO & 42.140 & 65.300 & 3.597 & 0.040 & 0.020 & 82.075 & 0.824 & 0.822 & 0.030 & 0.124 \\ 
  SparseStep & 4.360 & 41.900 & 0.017 & 0.030 & 0.015 & 84.450 & 0.847 & 0.844 & 0.148 & 0.108 \\ 
  Best Subset & 5.820 & 47.700 & 0.106 & 0.040 & 0.020 & 84.975 & 0.852 & 0.852 & 0.098 & 0.107 \\
  \hline
  SCAD & 24.680 & 66.300 & 1.823 & 0.030 & 0.020 & 86.700 & 0.867 & 0.870 & 0.035 & 0.093 \\ 
  MCP & 11.880 & 61.400 & 0.580 & 0.030 & 0.020 & 86.850 & 0.870 & 0.871 & 0.027 & 0.092 \\ 
  \hline
  SIS & 6.080 & 39.600 & 0.214 & 0.040 & 0.020 & 81.025 & 0.816 & 0.806 & 0.001 & 0.132 \\ 
  ISIS & 6.970 & 50.000 & 0.199 & 0.020 & 0.010 & 84.150 & 0.843 & 0.844 & 0.020 & 0.117 \\ 
  SIVS & 4.340 & 39.600 & 0.038 & 0.070 & 0.020 & 83.275 & 0.839 & 0.827 & 6.325 & 0.114 \\ 
  \hline
  varSelRF & 37.540 & 48.400 & 3.303 &  &  & 73.575 & 0.737 & 0.750 & 0.515 & 0.190 \\ 
  PIMP & 83.430 & 48.100 & 7.942 &  &  & 71.775 & 0.728 & 0.723 & 3.479 & 0.207 \\ 
  Boruta & 9.770 & 37.200 & 0.611 &  &  & 75.325 & 0.755 & 0.763 & 0.736 & 0.168 \\ 
  VSURF & 9.730 & 38.300 & 0.596 &  &  & 76.650 & 0.767 & 0.772 & 3.623 & 0.167 \\ 
  RRF & 3.900 & 26.200 & 0.129 &  &  & 75.950 & 0.764 & 0.761 & 0.075 & 0.169 \\ 
  NTA & 81.060 & 51.400 & 7.669 &  &  & 71.425 & 0.720 & 0.735 & 0.544 & 0.210 \\    
 \hline
  & \multicolumn{9}{c}{Correlated case} \\
  \hline
Method & Selected & Imp\% & Unimp\% & MSE & MAE & Accuracy & Precision & Recall & Time & Brier \\ 
  \hline
LASSO & 54.320 & 68.700 & 4.793 & 0.060 & 0.025 & 82.500 & 0.829 & 0.815 & 0.005 & 0.127 \\ 
  ElasticNet & 57.290 & 61.200 & 5.169 & 0.050 & 0.030 & 78.825 & 0.797 & 0.783 & 0.374 & 0.180 \\ 
  ALASSO & 41.280 & 64.400 & 3.519 & 0.040 & 0.020 & 82.325 & 0.828 & 0.817 & 0.023 & 0.128 \\ 
  SparseStep & 4.380 & 41.900 & 0.019 & 0.030 & 0.010 & 83.925 & 0.847 & 0.832 & 0.022 & 0.113 \\ 
  Best Subset & 5.740 & 48.000 & 0.095 & 0.040 & 0.020 & 84.375 & 0.853 & 0.835 & 0.019 & 0.109 \\ 
  \hline
  SCAD & 24.580 & 66.200 & 1.814 & 0.030 & 0.020 & 85.650 & 0.862 & 0.852 & 0.020 & 0.103 \\ 
  MCP & 11.760 & 60.000 & 0.582 & 0.040 & 0.020 & 85.400 & 0.861 & 0.848 & 0.014 & 0.102 \\ 
  \hline
  SIS & 4.990 & 36.300 & 0.137 & 0.050 & 0.020 & 80.925 & 0.818 & 0.800 & 0.003 & 0.138 \\ 
  ISIS & 6.940 & 49.100 & 0.205 & 0.020 & 0.020 & 83.800 & 0.842 & 0.833 & 0.020 & 0.122 \\ 
  SIVS & 4.240 & 39.100 & 0.033 & 0.060 & 0.020 & 83.200 & 0.838 & 0.826 & 6.063 & 0.120 \\ 
  \hline
  varSelRF & 37.500 & 48.700 & 3.296 &  &  & 74.400 & 0.756 & 0.731 & 0.446 & 0.189 \\ 
  PIMP & 86.970 & 47.800 & 8.302 &  &  & 71.375 & 0.737 & 0.693 & 3.014 & 0.206 \\ 
  Boruta & 11.060 & 36.800 & 0.745 &  &  & 75.025 & 0.755 & 0.743 & 0.908 & 0.173 \\ 
  VSURF & 9.690 & 35.800 & 0.617 &  &  & 74.450 & 0.750 & 0.737 & 5.053 & 0.175 \\ 
  RRF & 4.240 & 26.400 & 0.162 &  &  & 74.475 & 0.754 & 0.737 & 0.226 & 0.176 \\ 
  NTA & 84.070 & 51.800 & 7.969 &  &  & 72.075 & 0.743 & 0.696 & 0.480 & 0.205 \\ 
   \hline
\end{tabular}}
\end{table}

\begin{table}[hbt!]
\centering
\caption{performance metrics of the sixteen methods in the high dimensional simulation setting ($n = 500,m = 1000,p = 10$).}  
\label{tab:setup19}
\scalebox{0.8}{
\begin{tabular}{lcccccccccc}
   \hline
  & \multicolumn{9}{c}{Independent case} \\
  \hline
Method & Selected & Imp\% & Unimp\% & MSE & MAE & Accuracy & Precision & Recall & Time & Brier \\ 
  \hline
LASSO & 98.360 & 91.100 & 9.015 & 0.030 & 0.020 & 88.400 & 0.880 & 0.884 & 0.016 & 0.085 \\ 
  ElasticNet & 79.460 & 77.800 & 7.240 & 0.040 & 0.030 & 86.710 & 0.868 & 0.863 & 0.898 & 0.146 \\ 
  ALASSO & 48.740 & 82.600 & 4.089 & 0.020 & 0.020 & 89.060 & 0.886 & 0.890 & 0.052 & 0.077 \\ 
  SparseStep & 7.310 & 70.400 & 0.027 & 0.010 & 0.010 & 90.840 & 0.904 & 0.908 & 0.228 & 0.067 \\ 
  Best Subset & 8.160 & 74.600 & 0.071 & 0.010 & 0.010 & 90.930 & 0.904 & 0.911 & 0.267 & 0.066 \\ 
  \hline
  SCAD & 29.380 & 87.100 & 2.088 & 0.010 & 0.010 & 91.000 & 0.904 & 0.911 & 0.036 & 0.066 \\ 
  MCP & 14.950 & 82.800 & 0.674 & 0.010 & 0.010 & 91.230 & 0.909 & 0.911 & 0.053 & 0.064 \\ 
  \hline
  SIS & 8.080 & 63.900 & 0.171 & 0.020 & 0.010 & 88.850 & 0.884 & 0.887 & 0.002 & 0.078 \\ 
  ISIS & 14.990 & 80.400 & 0.702 & 0.040 & 0.020 & 88.890 & 0.885 & 0.890 & 0.084 & 0.091 \\ 
  SIVS & 4.350 & 42.900 & 0.006 & 0.050 & 0.020 & 84.730 & 0.841 & 0.850 & 14.583 & 0.104 \\ 
  \hline
  varSelRF & 28.590 & 62.500 & 2.257 &  &  & 81.680 & 0.814 & 0.815 & 1.541 & 0.154 \\ 
  PIMP & 89.880 & 61.300 & 8.460 &  &  & 78.680 & 0.788 & 0.784 & 10.999 & 0.186 \\ 
  Boruta & 8.430 & 51.900 & 0.327 &  &  & 82.440 & 0.821 & 0.824 & 0.859 & 0.132 \\ 
  VSURF & 9.580 & 56.100 & 0.401 &  &  & 82.560 & 0.827 & 0.818 & 8.952 & 0.132 \\ 
  RRF & 4.121 & 32.828 & 0.085 &  &  & 78.737 & 0.780 & 0.792 & 0.226 & 0.149 \\ 
  NTA & 73.660 & 64.800 & 6.786 &  &  & 79.180 & 0.789 & 0.793 & 1.450 & 0.182 \\ 
  \hline
  & \multicolumn{9}{c}{Correlated case} \\
  \hline
Method & Selected & Imp\% & Unimp\% & MSE & MAE & Accuracy & Precision & Recall & Time & Brier \\ 
  \hline
LASSO & 92.090 & 90.700 & 8.386 & 0.040 & 0.020 & 88.280 & 0.877 & 0.886 & 0.015 & 0.084 \\ 
  ElasticNet & 92.870 & 80.200 & 8.571 & 0.040 & 0.030 & 85.910 & 0.859 & 0.856 & 0.899 & 0.144 \\ 
  ALASSO & 47.060 & 83.400 & 3.911 & 0.020 & 0.020 & 89.130 & 0.884 & 0.896 & 0.053 & 0.077 \\ 
  SparseStep & 7.360 & 71.300 & 0.023 & 0.010 & 0.010 & 90.380 & 0.901 & 0.903 & 0.014 & 0.068 \\ 
  Best Subset & 7.790 & 73.500 & 0.044 & 0.010 & 0.010 & 90.730 & 0.903 & 0.909 & 0.029 & 0.066 \\ 
  \hline
  SCAD & 26.260 & 86.700 & 1.777 & 0.010 & 0.010 & 90.810 & 0.903 & 0.912 & 0.030 & 0.066 \\ 
  MCP & 13.800 & 81.700 & 0.569 & 0.010 & 0.010 & 90.800 & 0.902 & 0.912 & 0.039 & 0.065 \\ 
  \hline
  SIS & 6.940 & 61.400 & 0.081 & 0.020 & 0.010 & 88.910 & 0.887 & 0.888 & 0.006 & 0.079 \\ 
  ISIS & 14.640 & 80.600 & 0.665 & 0.070 & 0.030 & 88.850 & 0.886 & 0.887 & 0.139 & 0.095 \\ 
  SIVS & 4.360 & 43.400 & 0.002 & 0.050 & 0.020 & 85.210 & 0.846 & 0.856 & 13.534 & 0.104 \\ 
  \hline
  varSelRF & 30.810 & 61.100 & 2.495 &  &  & 81.460 & 0.813 & 0.814 & 1.441 & 0.151 \\ 
  PIMP & 91.290 & 61.900 & 8.596 &  &  & 79.200 & 0.793 & 0.794 & 10.411 & 0.180 \\ 
  Boruta & 12.050 & 54.800 & 0.664 &  &  & 82.120 & 0.819 & 0.821 & 1.946 & 0.135 \\ 
  VSURF & 9.540 & 55.800 & 0.400 &  &  & 82.460 & 0.819 & 0.829 & 8.680 & 0.131 \\ 
  RRF & 4.260 & 32.000 & 0.107 &  &  & 78.760 & 0.782 & 0.790 & 0.210 & 0.147 \\ 
  NTA & 80.210 & 66.900 & 7.427 &  &  & 79.550 & 0.794 & 0.799 & 1.536 & 0.178 \\ 
   \hline
\end{tabular}}
\end{table}

\begin{figure}[!ht]
\centering 
\includegraphics[width = \textwidth]{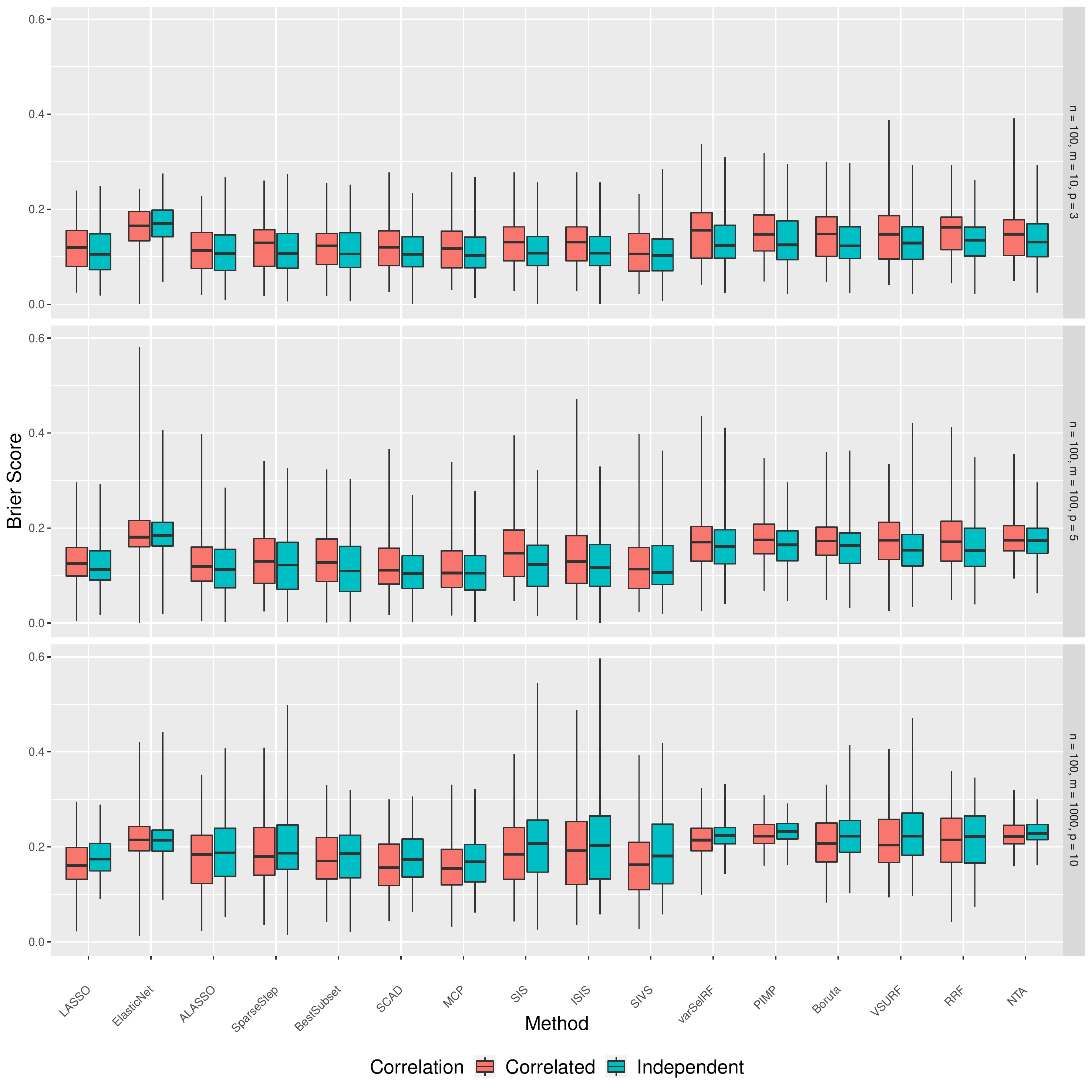}
\caption{Boxplots of the prediction Brier scores across various iterations in case of $n = 100$ for different
dimensions.}
\label{fig:brier1}
\end{figure}  

\begin{figure}[!ht]
\centering 
\includegraphics[width = \textwidth]{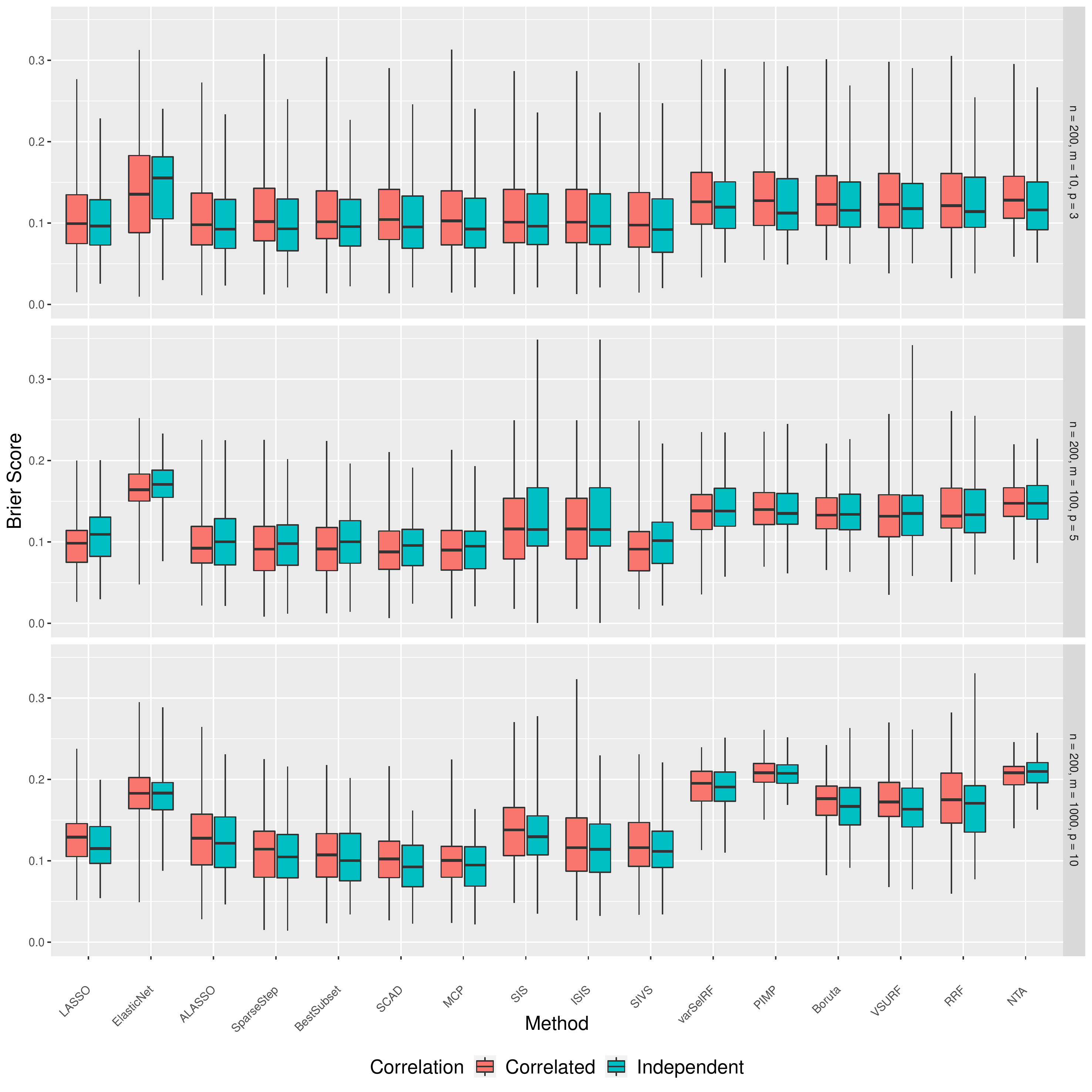}
\caption{Boxplots of the prediction Brier scores across various iterations in case of $n = 200$ for different
dimensions.}
\label{fig:brier2}
\end{figure}  

\begin{figure}[!ht]
\centering 
\includegraphics[width = \textwidth]{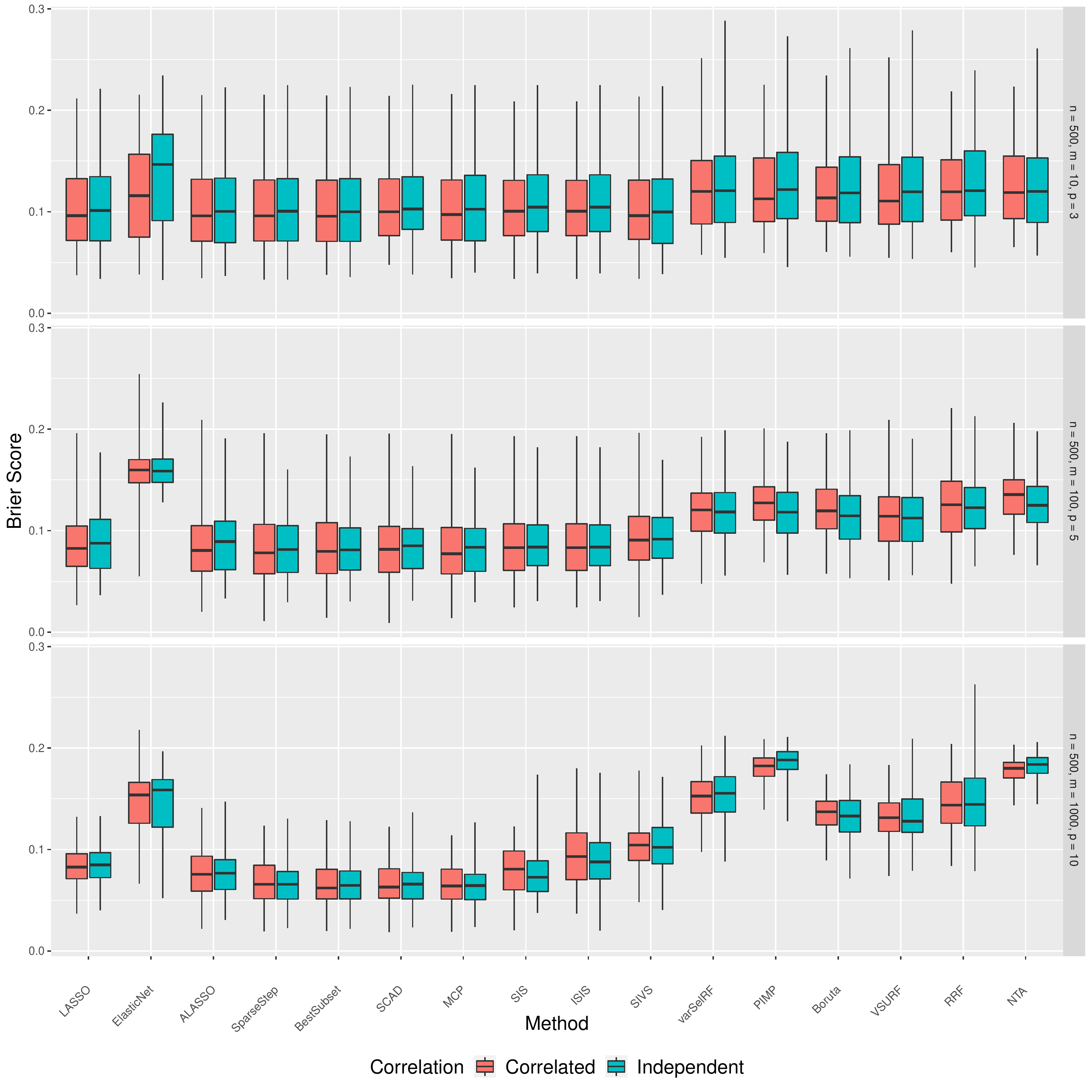}
\caption{Boxplots of the prediction Brier scores across various iterations in case of $n = 500$ for different
dimensions.}
\label{fig:brier}
\end{figure}

\end{document}